\documentclass[lettersize,journal]{IEEEtran}

\usepackage[T1]{fontenc}
\usepackage{amsmath}
\usepackage{seqsplit} %to split long words automatically
\usepackage[normalem]{ulem} %to remove underline from title bibliography
\usepackage{soul}
\usepackage{stfloats}
\usepackage[pdftex]{graphicx}
\usepackage{algorithm,algorithmic}
\usepackage{listings}
\usepackage{tikz}
\usepackage{array}
\newcolumntype{x}[1]{>{\centering\arraybackslash\hspace{0pt}}m{#1}}
\newcolumntype{j}[1]{>{\raggedright\arraybackslash\hspace{0pt}}m{#1}}
\usepackage{pdfrender}
\usepackage{multirow}
\usepackage{svg}
\usepackage{graphicx}
\usepackage{url}

\usepackage{balance}
\usepackage{listings}
\usepackage{subcaption}

\usepackage[capitalise, noabbrev]{cleveref}
\usepackage{listings}
\usepackage{xcolor}
\usepackage[square,numbers]{natbib}
\usepackage{enumitem}

\usepackage{booktabs}

\usepackage{xurl} 
\usepackage[nolist]{acronym}
\usepackage{msc}

\usepackage{mathptmx} % Sets Times New Roman as the font
\usepackage{comment}

\begin{document}

\begin{acronym}
    \acro{5GAA}{5G Automotive Associations}
    \acro{ADAS}{Advanced Driver-Assistance Systems}
    \acro{AAOS}{Android Automotive OS}
    \acro{ARP}{Address Resolution Protocol}
    \acro{AI}{Artificial Intelligence}
    \acro{AitD}{Attacker-in-the-Device}
    \acro{API}{Application Programming Interface}
    \acro{ASU}{Arizona State University}
    \acro{AV}{Autonomous Vehicle}
    \acro{CAN}{Controller Area Network}
    \acro{CI/CD}{Continuous Integration/Continuous Deployment}
    \acro{CNN}{Convolutional Neural Network}
    \acro{CV}{Connected Vehicle}
    \acro{CV2X}{Cellular Vehicle-to-Everything}
    \acro{DY}{Dolev-Yao}
    \acro{DoS}{Denial of Service}
    \acro{DFS}{Depth-First Search}
    \acro{DL}{Deep Learning}
    \acro{DSRC}{Dedicated Short-Range Communication}
    \acro{DTW}{Dynamic Time Warping}
    \acro{E/E}{Electrical/Electronic}
    \acro{ECC}{Elliptic Curve Cryptography}
    \acro{ECU}{Electronic Control Unit}
    \acro{ERV}{Emergency Response Vehicle}
    \acro{EV}{Electric Vehicle}
    \acro{GDPR}{General Data Protection Regulation}
    \acro{GRU}{Graphics Processing Unit}
    \acro{GPU}{Gated Recurrent Unit}
    \acro{HOV}{High-Occupancy Vehicle}
    \acro{IDS}{Intrusion Detection System}
    \acro{IDPS}{Intrusion Detection and Prevention System}
    \acro{IMU}{Inertial Measurement Unit}
    \acro{IoT}{Internet of Things}
    \acro{IPS}{Intrusion Prevention System}
    \acro{IR}{Infrared}
    \acro{ITS}{Intelligent Transportation System}
    \acro{IVI}{In-Vehicle Infotainment}
    \acro{IVN}{In-Vehicle Network}
    \acro{LED}{Light-Emitting Diode}
    \acro{LIN}{Local Interconnect Network} 
    \acro{LSTM}{Long Short-Term Memory}
    \acro{LOS}{Line-of-Sight}
    \acro{LR}{Literature Review}
    \acro{MAC}{Message Authentication Codes}
    \acro{MIT}{Massachusetts Institute of Technology}
    \acro{MFA}{Multi-Factor Authentication}
    \acro{ML}{Machine Learning}
    \acro{MQTT}{Message Queuing Telemetry Transport}
    \acro{NLOS}{Non-Line-of-Sight}
    \acro{OBD-II}{On-Board Diagnostics-II}
    \acro{OBE}{On-Board Equipment}
    \acro{OEM}{Original Equipment Manufacturer}
    \acro{OCC}{Optical Camera Communication}
    \acro{OOK}{On-Off Keying} 
    \acro{OTA}{Over-the-Air}
    \acro{PII}{Personally Identifiable Information}
    \acro{PIN}{Personal Identification Number}
    \acro{PKI}{Public Key Infrastructure}
    \acro{PWM}{Pulse Width Modulation}
    \acro{RA}{Registration Authority}
    \acro{ReLU}{Rectified Linear Unit}
    \acro{RF}{Radio Frequency}
    \acro{RGB}{Red, Green, and Blue}
    \acro{RQ}{Research Question}
    \acro{ROS}{Robot Operating System}
    \acro{RSE}{Roadside Equipment}
    \acro{RSU}{Roadside Unit}
    \acro{SaaS}{Software as a Service}
    \acro{SCMS}{Security Credential Management System}
    \acro{SoF}{Start of Frame}
    \acro{SVSE}{Secure Vehicle Software Engineering}
    \acro{SDN}{Software-Defined Networking}
    \acro{SDV}{Software-Defined Vehicle}
    \acro{SecOC}{Secure Onboard Communication}
    \acro{SOA}{Service Oriented Architecture}
    \acro{S-ARP}{Secure ARP}
    \acro{TAK}{Title-Abstract-Keywords}
    \acro{TESLA}{Timed Efficient Stream Loss-Tolerant Authentication}
    \acro{TLS}{Transport Layer Security}
    \acro{ToS}{Terms of Service}
    \acro{UAV}{Unmanned Aerial Vehicle}
    \acro{V2I}{Vehicle-to-Infrastructure}
    \acro{V2N}{Vehicle-to-Network}
    \acro{V2X}{Vehicle-to-Everything}
    \acro{V2V}{Vehicle-to-Vehicle}
    \acro{VANET}{Vehicular Ad Hoc Networks}
    \acro{VLAN}{Virtual LAN}
    \acro{VLC}{Visible Light Communication}
    \acro{YOLO}{You Only Look Once}
    \acro{WAVE}{Wireless Access in Vehicular Environments}
\end{acronym}

\title{Vehicular Communication Security: Multi-Channel and Multi-Factor Authentication}

\author{Marco~De~Vincenzi, Shuyang Sun, Chen Bo Calvin Zhang, Manuel Garcia, Shaozu Ding, \\ Chiara Bodei, Ilaria Matteucci, Sanjay E. Sarma, and Dajiang~Suo% <-this % stops a space

\thanks{M. De Vincenzi was with the AUTO-ID Lab, Massachusetts Institute of Technology (MIT) and the Polytechnic School of the Ira A. Fulton Schools of Engineering, Arizona State University; and he is with the Institute of Informatics and Telematics, 57124, Pisa, Italy; email: marco.devincenzi@iit.cnr.it.} 

\thanks{Shuyang Sun is with the Department of Engineering Science, University of Oxford, OX1 3LZ Oxford, U.K. Email: kevinsun@robots.ox.ac.uk.}

\thanks{Chen Bo Calvin Zhang was with the Department of Computer Science, ETH Zurich, 8092 Zurich, Switzerland; and with the Department of Electrical Engineering and Computer Science, Massachusetts Institute of Technology (MIT), USA. Emails: zhangca@ethz.ch, cbczhang@mit.edu.}
\thanks{C. Bodei is with the Department of Computer Science, Università di Pisa, Pisa, Italy. Email: chiara.bodei@unipi.it.}
\thanks{I. Matteucci is with the Institute of Informatics and Telematics, 56124 Pisa, Italy. Email: ilaria.matteucci@iit.cnr.it.} 
\thanks{M. Garcia and S. Ding are with the Polytechnic School of the Ira A. Fulton Schools of Engineering, Arizona State University, Tempe, AZ 85281, USA. Emails: mgarci84@asu.edu, sding32@asu.edu.}

\thanks{Sanjay E. Sarma is with the Department of Mechanical Engineering, Massachusetts Institute of Technology, Cambridge, MA 02139, USA. Email: sesarma@mit.edu.}

\thanks{D. Suo is with the Polytechnic School of the Ira A. Fulton Schools of Engineering, Arizona State University, Tempe, AZ 85281, USA. Email: dajiang.suo@asu.edu (Corresponding author).}
}

%The paper headers

 %Remember, if you use this you must call \IEEEpubidadjcol in the second

\maketitle
\thispagestyle{empty}
\pagestyle{empty}

%%%%%%%%%%%%%%%%%%%%%%%%%%%%%%%%%%%%%%%%%%%%%%%%%%%%%%%%%%%%%%%%%%%%%%%%%%%%%%%%
\begin{abstract}
Secure and reliable communications are crucial for Intelligent Transportation Systems (ITSs), where Vehicle-to-Infrastructure (V2I) communication plays a key role in enabling mobility-enhancing and safety-critical services. Current V2I authentication relies on credential-based methods over wireless Non-Line-of-Sight (NLOS) channels, leaving them exposed to remote impersonation and proximity attacks. To mitigate these risks, we propose a unified Multi-Channel, Multi-Factor Authentication (MFA) scheme that combines NLOS cryptographic credentials with a Line-of-Sight (LOS) visual channel. Our approach leverages a challenge–response security paradigm: the infrastructure issues ``challenges'' and the vehicle’s headlights respond by flashing a structured sequence containing encoded security data. Deep learning models on the infrastructure side then decode the embedded information to authenticate the vehicle. Real-world experimental evaluations demonstrate high test accuracy, reaching an average of 95\% and 96.6\%, respectively, under various lighting, weather, speed, and distance conditions. Additionally, we conducted extensive experiments on three state‐of‐the‐art deep learning models, including detailed ablation studies for decoding flashing sequence. Our results indicate that the optimal architecture employs a dual‐channel design, enabling simultaneous decoding of the flashing sequence and extraction of vehicle spatial and locational features for robust authentication.
\end{abstract}


%%%%%%%%%%%%%%%%%%%%%%%%%%%%%%%%%%%%%%%%%%%%%%%%%%%%%%%%%%%%%%%%%%%%%%%%%%%%%%%%
\section{Introduction} \label{sec:introduction}
\IEEEPARstart{T}{he} future of transportation is not only green or autonomous, it is also connected, intelligent, and increasingly vulnerable to cyber threats \cite{sdv, devincenzi2024contextualizingsecurityprivacysoftwaredefined}. As vehicles continuously exchange critical information with their surroundings, establishing secure and trustworthy communication becomes essential for both mobility and safety.
Examples include \ac{V2V} communication, where nearby vehicles share data to coordinate movement and avoid collisions \cite{8463512}, and \ac{V2I} communication, where vehicles interact with the road infrastructure to request and receive traffic signal timings \cite{10.1117/12.3013514}.
Focusing on \ac{V2I} applications, we consider a scenario in which vehicles must authenticate themselves to access restricted lanes or road segments, such as high-occupancy vehicle lanes, airport zones or military areas, or, more generally, receive services. This use case typically involves secure exchanges with the road infrastructure to verify the vehicle's identity and authorization status before granting access to the controlled area or enabling a specific service. The vehicle authentication process, which involves proving the identity of the vehicle, is crucial to ensuring both safety and security \cite{Abu-Nimeh2011, 10818588}. 

\subsection{Motivations}

Vulnerabilities in authentication processes arise due to the inherent characteristics of wireless communication channels. In particular, adversaries can exploit them to impersonate vehicles, manipulate signal priorities, or inject false information into the network. These attacks can compromise the integrity and security of services and restricted zones, posing significant safety risks \citep{Dasgupta2022, irfan2022reinforcementlearningbasedcyberattack, 9916277}. Current \ac{V2I} authentication schemes, which predominantly focus on credential verification rather than confirming the authenticity and trustworthiness of entities, fail to mitigate these threats \citep{V2I1, surveyV2Xattack2, surveyV2Xattack3}.
Recently, researchers explored the use of side communication channels, particularly \ac{LOS} channels, to exchange authentication data between vehicles and increase security \cite{visual1}.

\subsection{Our contribution}

We propose a unified Multi-Channel and \ac{MFA} scheme that introduces a comprehensive and novel authentication framework, which we also validate through experimental testing. Our scheme fully integrates \ac{LOS} communication and its corresponding challenge-response mechanism. The multi-channel approach ensures that the communicating entity is a real vehicle, as visual confirmation provides an additional layer of authentication based on physical presence. Finally, we assess its robustness through a comprehensive real-world evaluation using a SlowFast network model to detect vehicle response.

Our key contributions are the following.
\begin{itemize}
    \item \textit{Designed Security Scheme:} 
    We introduce a novel security scheme that strengthens vehicle authentication in V2I communication scenarios, addressing threats from both remote and physically present attackers.
    \item \textit{Realization and Evaluation of the \ac{LOS} Mechanism:} The \ac{LOS} channel is implemented using visible light, where the vehicle flashes its headlights in response to a challenge generated by the road infrastructure. We validated this mechanism through hardware-in-the-loop experiments, starting with an RC-car and subsequently in a real-world vehicle environment. 
    \item \textit{Computer vision model:} We conducted an analysis to guide the design of the neural network architecture for visual channel-based authentication. The flashing sequence is classified using a computer vision system based on SlowFast, a two-stream action recognition architecture consisting of two \acp{CNN} networks. This approach offers high classification accuracy and strong generalization capabilities. 
\end{itemize}
The results demonstrated the feasibility of using the proposed scheme for vehicle authentication in real-world scenarios, taking into account variables such as different lighting conditions, environments, weather, speeds, distances, and even the presence of other vehicles.

\subsection{Organization of the Paper} \label{sec:organization}

The paper is organized as follows. In \cref{sec:rw}, we review related work on \ac{V2I} authentication, with a focus on \ac{MFA} methods using visual channels. \Cref{sec:reference_scenarios} defines the attack model and threat scenarios motivating a robust \ac{V2I} authentication scheme. In \cref{sec:architecture}, we present our scheme, detailing its components, including the \ac{LOS}-enabled challenge-response mechanism and the security frame. \Cref{sec:secanalysis} analyzes the security properties of the scheme and how it addresses key threats. \Cref{sec:setup} outlines the experimental implementation, including hardware, data collection in two testbeds, and the deep learning-based vision model. \Cref{sec:MLresults} and \cref{sec:MLdiscussion} analyze the experimental results and assess performance. Finally, \cref{sec:conclusion} summarizes the work and discusses future directions.

\section{Related work} \label{sec:rw}

\ac{MFA} authentication in \ac{V2I} communication has gained significant research interest in the past decade. Previous work explores \ac{MFA} schemes for \ac{V2I} contexts \cite{10.1145/3605098.3636102}, in addition to vehicle light recognition for autonomous systems \cite{taillight}. The following studies provide key foundations that inform our work.

Our work is based on the method of \citet{dajiangMFA}, which counteracts impersonation and message fabrication by using \ac{LOS} communication as a second factor. Their approach requires vehicles to respond to a \ac{NLOS} challenge through a directional \ac{LOS} channel, making the impersonation of stationary adversaries difficult. A core innovation is the use of vehicle movement for validation, enforcing physical constraints to confirm legitimacy. In contrast, we eliminate LOS bottlenecks by using lightweight visual patterns and avoid reliance on infrared LEDs or custom hardware. Using native vehicle visuals, we improve practicality and reduce costs. Furthermore, our real-world tests with the SlowFast network address open challenges such as timing accuracy and authentication at varying speeds.

Based on this, \citet{benDw} propose an MFA approach using QR codes transmitted via the \ac{LOS} channel and recognized by infrastructure-side cameras using neural networks \citep{redmon2016you}. However, their method requires front-facing displays in vehicles, raising cost concerns. Our system instead uses existing visual elements and custom patterns to reduce hardware dependencies.

\citet{visual1} present a vision-based MFA and localization scheme for \ac{AV}s, using vehicle headlights and cameras onboard to transmit and verify nonce-based authentication messages. This approach inspired both \citet{dajiangMFA} and our own scheme, which we improved and further validated in dynamic real-world scenarios.

\citet{Arfaoui} survey physical layer security techniques in \ac{VLC}, highlighting the advantages of \ac{LOS}-based channels for confidentiality and authentication. Although theoretical and system-agnostic, their principles of spatial modulation, artificial noise, and beamforming influence our practical use of light as a secure side channel.

\citet{Singh} introduce a hybrid V-VLC/V-\ac{RF} model for intersections, dynamically switching channels based on quality. Their method improves outage and delay metrics through stochastic analysis, but is based on dual transceivers and complex switching. Our design, on the contrary, uses a fixed visual challenge-response without additional hardware, enabling lightweight authentication in dynamic conditions.

\citet{Rowan} propose a secure \ac{V2V} framework using VLC, acoustic channels, and blockchain PKI for secondary key exchange. Although resistant to RF jamming, their method targets low-throughput \ac{V2V} contexts and requires blockchain-based key management. Our work avoids such complexity, achieving secure \ac{V2I} authentication through native light signals and a streamlined scheme.

\citet{Shaaban} examine VLC secrecy in platooning via LED semiangle tuning and spatial zone definition to reduce eavesdropping. Their strategy, while effective in static formations, demands fine calibration. Our solution provides robustness in dynamic \ac{V2I} settings without tuning, offering flexible deployment through visual sensing and existing lights.

\section{Attack Model} \label{sec:reference_scenarios}

\begin{figure*}[t!]
    \centering
    \begin{minipage}{0.245\textwidth}
        \includegraphics[width=\linewidth]{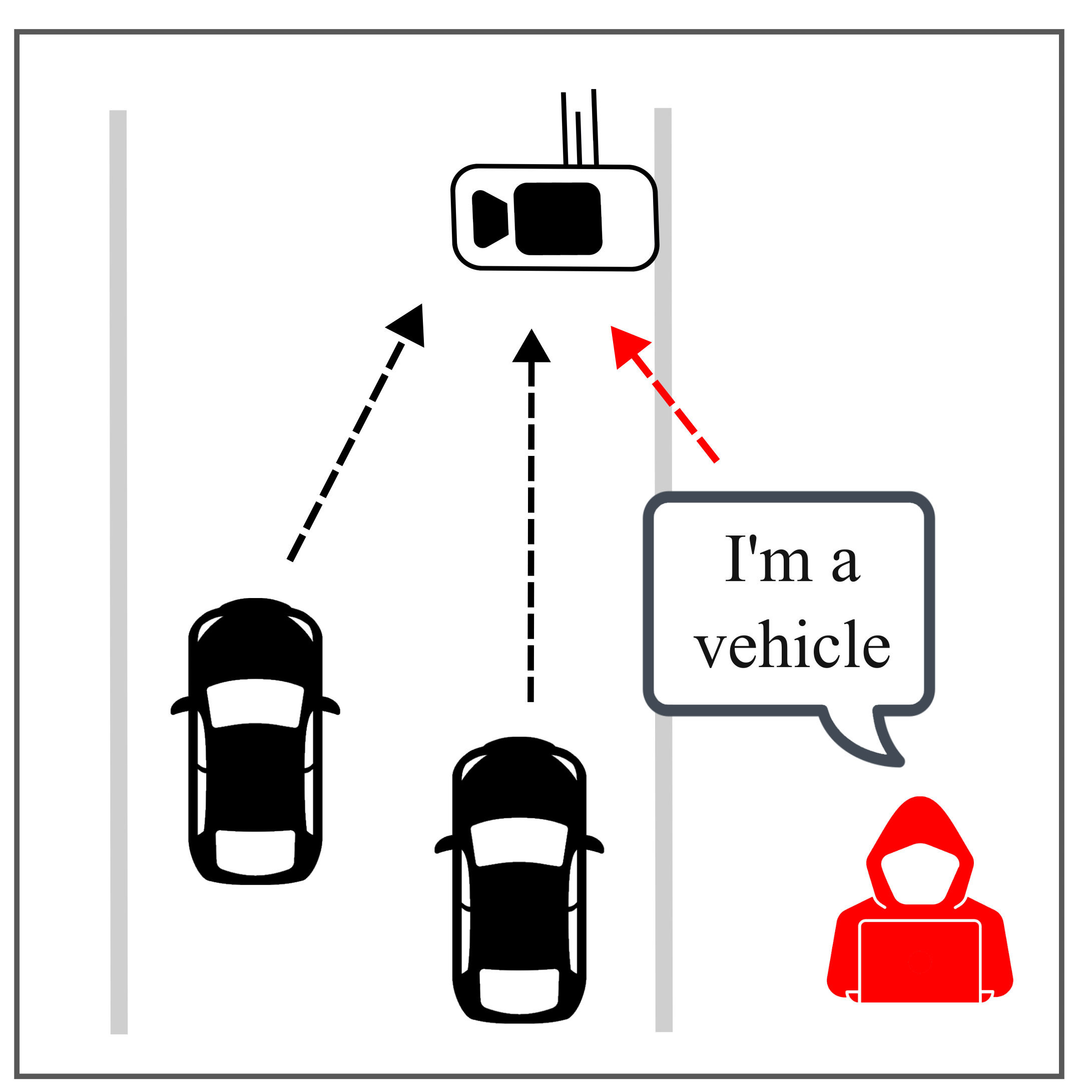}
        \centering (1) \ac{V2I} Communications.
        \label{fig:scenario1}
    \end{minipage}
    \begin{minipage}{0.245\textwidth}
        \includegraphics[width=\linewidth]{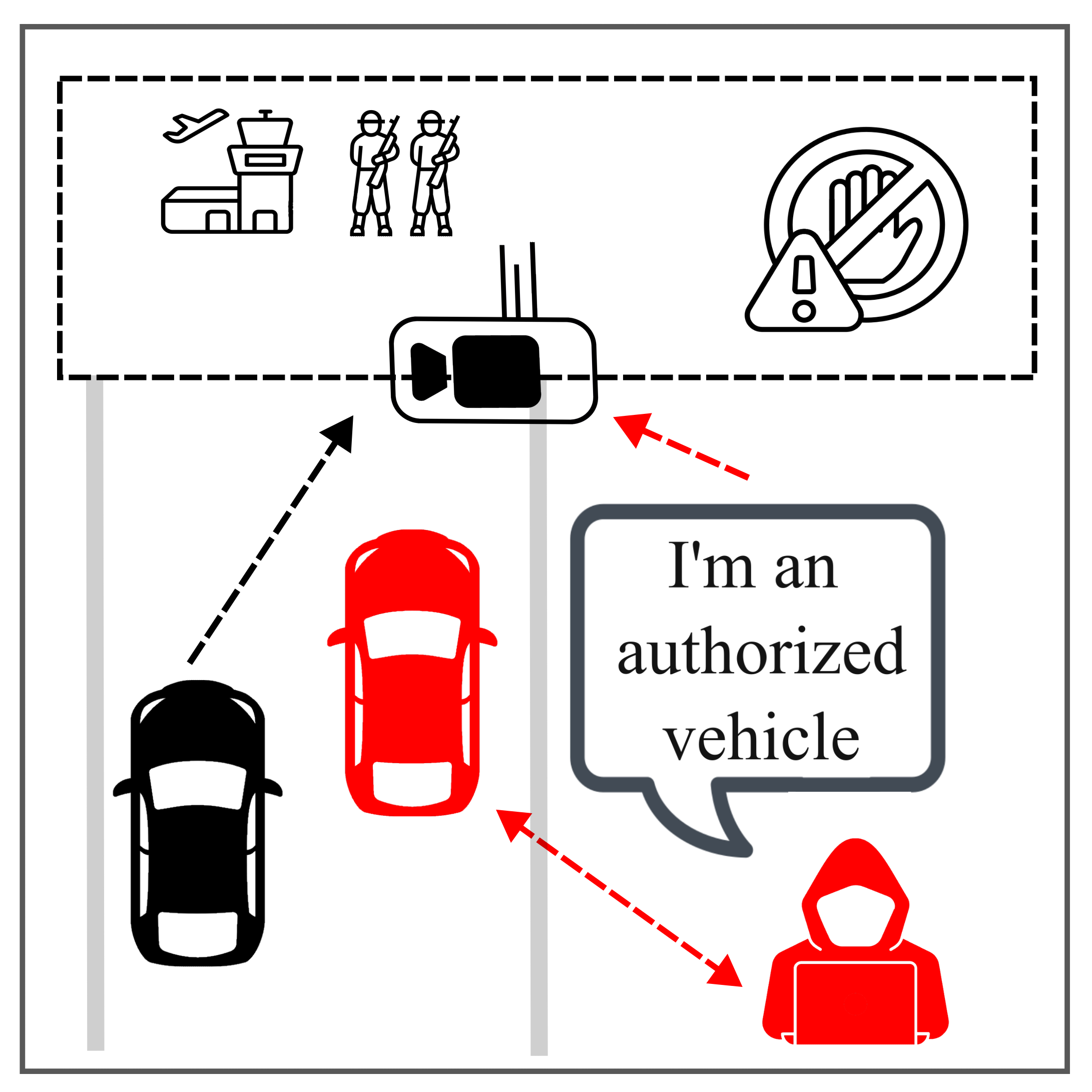}
        \centering (2) Restricted Area.
        \label{fig:scenario2}
    \end{minipage}
    \begin{minipage}{0.245\textwidth}
        \includegraphics[width=\linewidth]{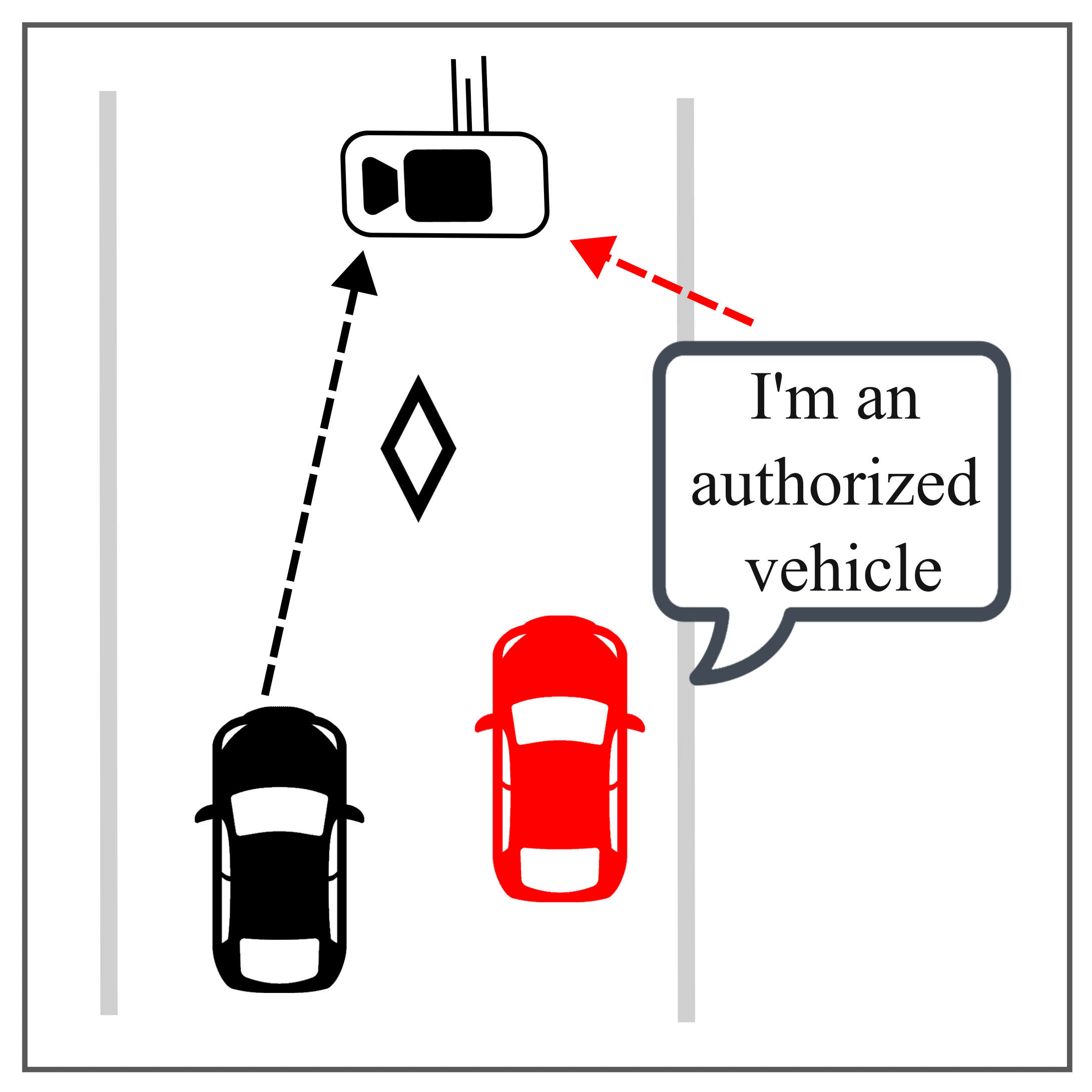}
        \centering (3) Reserved Lane.
        \label{fig:scenario3}
    \end{minipage}
    \begin{minipage}{0.245\textwidth}
        \includegraphics[width=\linewidth]{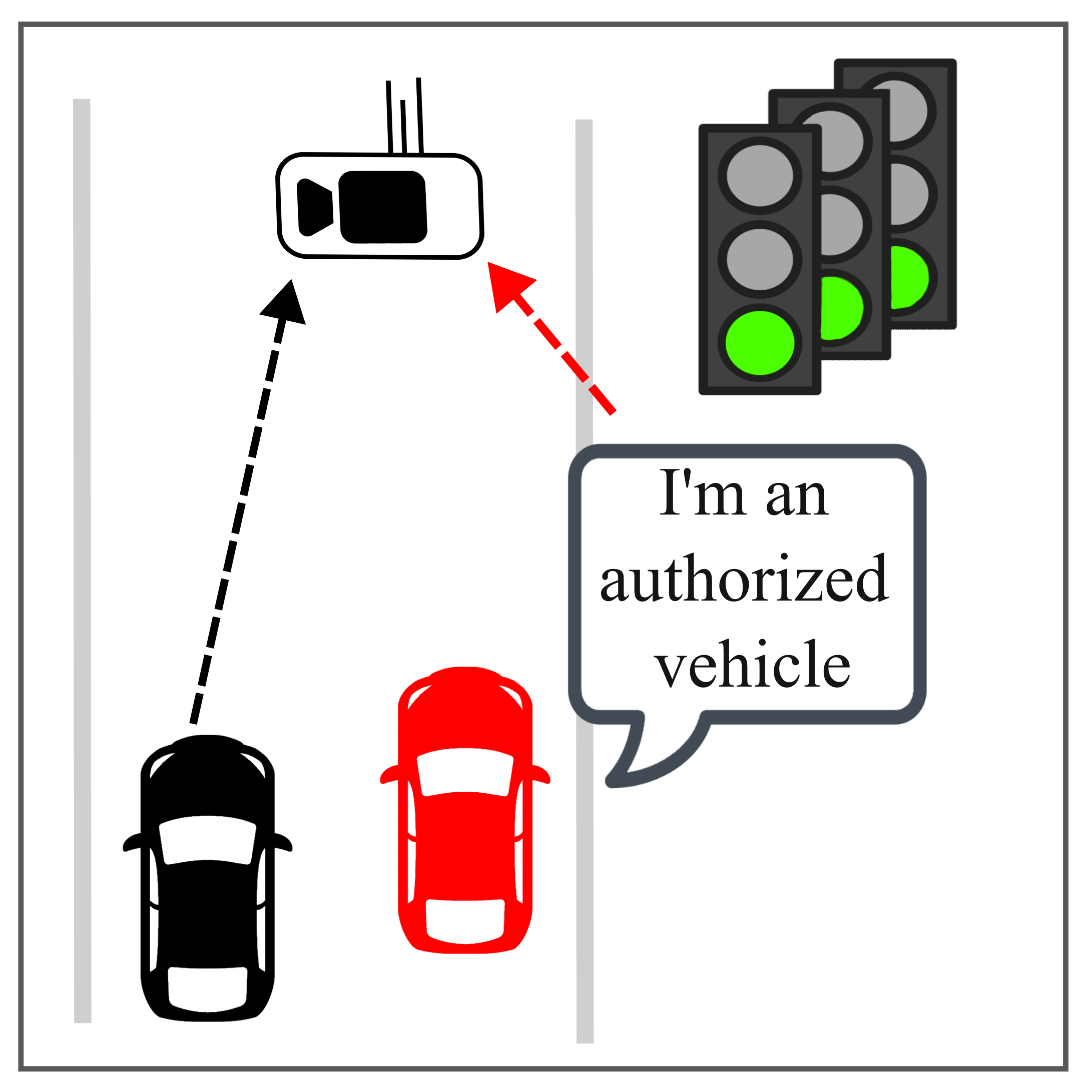}
        \centering \small{(4) Signal Preemption.}
        \label{fig:scenario4}
    \end{minipage}
    \caption{Critical \ac{ITS} scenarios requiring authentication mechanisms, 
    %to mitigate safety and security risks, 
    with attacker devices and channels controlled by attackers highlighted in red.}
    \label{fig:its_scenarios}
\end{figure*}

Our proposed scheme focuses on vehicle authentication in the four critical scenarios shown in \cref{fig:its_scenarios},
where authentication failures could lead to operational, security, and safety risks.
These scenarios, based on the guidelines provided by \ac{5GAA} \citep{5gaaUS}, include standard \ac{V2I} communications, in which vehicles interact with the road
infrastructure (\cref{fig:its_scenarios}.1), access to restricted areas (\cref{fig:its_scenarios}.2), use of reserved lanes (\cref{fig:its_scenarios}.3), and
signal preemption, as an example of required service, used to override normal traffic light operation and allows vehicles to bypass traffic (\cref{fig:its_scenarios}.4). 
%They are generally done to assist emergency vehicles.
In all of these cases, authentication is critical to prevent unauthorized impersonation and potential attacks. 

To explore potential threat scenarios, we use the STRIDE framework \citep{stride}, which classifies threats according to their nature and impact. STRIDE categorizes security threats into six types: Spoofing (S), Tampering (T), Repudiation (R), Information Disclosure (I), Denial of Service (DoS), and Elevation of Privilege (E). In addition, it provides a structured methodology for identifying the most critical risks and determining appropriate mitigation strategies.

Moreover, we adopt the \ac{DY} model \citep{dolevYao} to define the capabilities of an adversary in our security analysis. This model assumes a powerful attacker who can intercept, modify, and inject messages within the network.
Unlike traditional computer networks, vehicle communication infrastructure introduces a peculiar threat landscape in which attackers can operate remotely or in close physical proximity to critical infrastructure such as \acp{RSU} or cameras. This dual attack modality, where adversaries can exploit cryptographic weaknesses from a distance or physically interfere with communication channels, poses distinct challenges to authentication and security.

Based on these attacker capabilities, our analysis identifies the following three key threats that must be taken into account. In parentheses, the STRIDE \cite{stride} categories of threats involved.

\begin{itemize}
\item 
{\it Remote Impersonation Attacks (S, T):} 
Attackers remotely impersonate vehicles by injecting falsified messages into the network, exploiting weaknesses in the authentication scheme. For example, in (\cref{fig:its_scenarios}.1), attackers can disrupt operations by intercepting or modifying messages, leading to safety and operational risks. 

\item {\it Proximity-Based Attacks (S, E, T):}
Proximity-based attacks occur when an attacker or a vehicle operates near critical infrastructure, such as an \ac{RSU} or surveillance system, to impersonate an authorized vehicle. This type of attack is particularly critical in scenarios that require access to restricted areas, such as military bases or airports (\cref{fig:its_scenarios}.2), and the use of a reserved lane (\cref{fig:its_scenarios}.3), such as bus or \ac{HOV} lanes.
By exploiting authentication weaknesses, attackers can impersonate legitimate vehicles, gaining unauthorized access to restricted areas or reserved lanes. These attacks disrupt operational efficiency, compromise security, and undermine trust in the system.

\item
{\it Traffic Signal Preemption Attacks (S,T,D)}: In scenarios such as those shown in (\cref{fig:its_scenarios}.4), these attacks occur when malicious entities spoof vehicle identities to manipulate traffic signals, gain unauthorized priority access, and disrupt normal traffic operations (see, e.g., \cite{cybernews2024dutchtrafficlights}).
These attacks can compromise the effectiveness of emergency response, create security risks, and cause widespread traffic congestion. In addition, they may interfere with green wave systems, further amplifying traffic disruptions.

\end{itemize}

\section{Multi-Channel Multi-Factor Scheme} \label{sec:architecture} 
\begin{figure}[hbt!]
    \centering
    \includegraphics[width=\linewidth]{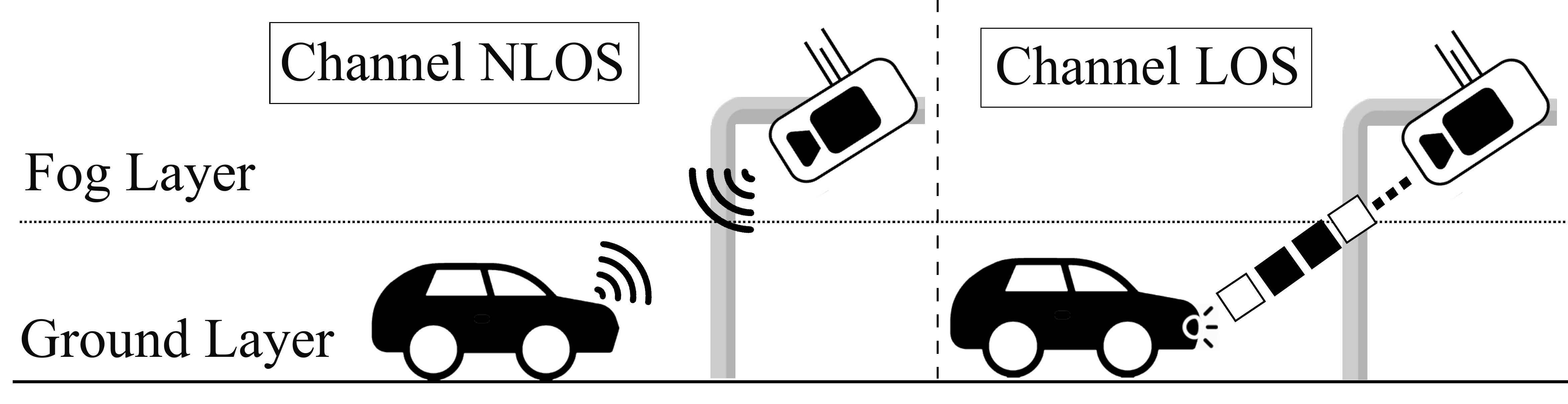}
    \caption{Two-channels (\ac{NLOS} and \ac{LOS}) authentication process.}
    \label{fig:scenario}
\end{figure}

Our proposed authentication process introduces an authentication process with a fully implemented \ac{LOS} communication and response mechanism. 
The scheme integrates multiple factors across \ac{NLOS} and \ac{LOS} channels as shown in \cref{fig:scenario} 
%As described in \cref{{fig:sequence_diagram}}, 
and consists of the following three phases.%: Setup, Challenge (chall), and Check, covering eight messages.
\begin{itemize}
    \item {\it \ac{NLOS} Phase}:
    the vehicle uses the \ac{NLOS} channel to transmit its unique security credential to a \ac{RA} over a \ac{TLS}-protected channel, as defined in IEEE 1609.2.1 \cite{ieee16092}.
    This serves as the first authentication factor (\textit{something you know}). %in a multi-factor authentication process.
    For example, the exchange of security credentials can be carried out with the support of \ac{PKI} \cite{10075082, brecht2018security}. 
    \item {\it \ac{LOS} Phase}: After verifying the first authentication factor, the infrastructure requests a second one (e.g., \textit{something you are} or \textit{something you know}). The \ac{RSU} then sends the vehicle a randomized challenge, which the vehicle answers through the optical channel.
    \item {\it Check Phase}: The \ac{RA} verifies the correctness of the response. Upon successful validation, an authentication token is issued to the vehicle.

\end{itemize}

\subsection{Challenge-Response mechanism
} \label{sec:challengeresponse}

A key novelty of the proposed scheme is the implementation of the challenge-response mechanism. This involves generating \textit{security frames}, structured sequences that the vehicle must follow to correctly respond to the challenge of the infrastructure. These frames define the response format within the challenge-response mechanism. As the response is sent over the LOS channel, the frames must be physically encoded. Various methods are possible; in our case, the vehicle's headlights reproduce the frame patterns by flashing (\cref{sec:occ}). Since headlights are standard equipment, no additional hardware is needed. The flashed response is then captured by visual sensors, such as cameras, at the other end of the \ac{LOS} channel.

\begin{figure}[t!]
    \centering
    \includegraphics[width=0.90\linewidth]{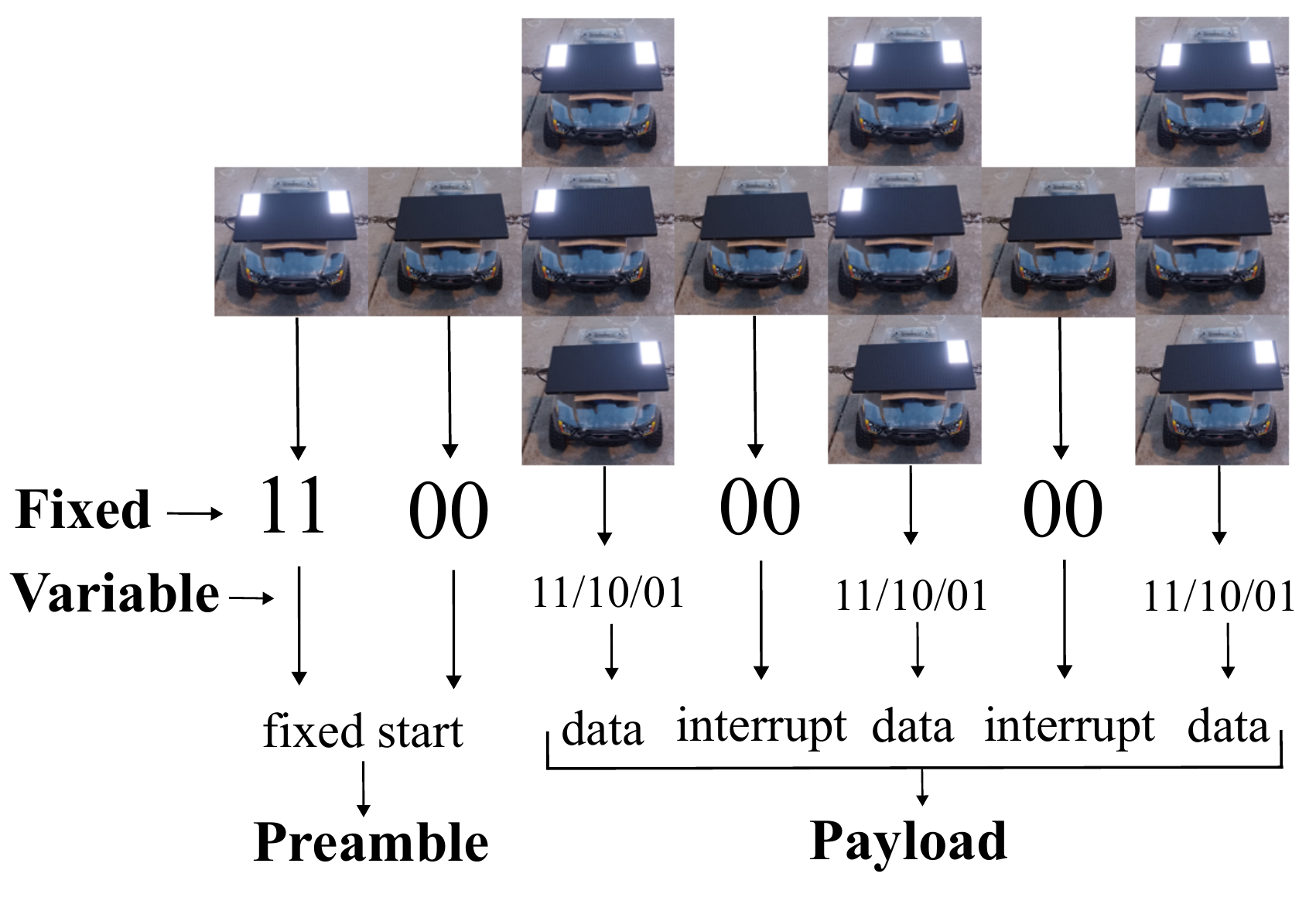}
    \caption{The 14-bit security frame and its components.}
    \label{fig:frame}
\end{figure}

The vehicle's headlights naturally encode the binary values of the security frames: a headlight turned on represents bit 1, and a headlight turned off represents bit 0.
As a result, using both headlights in a single flash allows the transmission of four possible values: 11, 10, 01, or 00.
As illustrated in \cref{fig:frame}, in our implementation,
the security frame is composed of 14 bits and divided into two parts.
\begin{itemize}
\item {\it Preamble}: is composed of the sequence $11-00$, where $11$ represents the start of the message and $00$ is the interrupt sequence. 
Notifies the camera when to begin to detect the payload.
\item {\it Payload}: this is the actual data field and is composed of 5 flashes: 3 containing the information (different from 00) and 2 interrupts (00). 
\end{itemize}
 
Therefore, the possible sequences are 27 ($3^3$) in total.
The length of the frame and of its parts has been obtained as a trade-off between security constraints and timing for authentication, in order to finish the challenge-response process before the vehicle moves out of the coverage area by the camera.

\subsubsection*{Choice of the length of the security frame}
To determine an acceptable length, we derived an
equation that relates the bit length n, the time available for
authentication, vehicle speed and computation and transmission time.

\begin{itemize}
\item 
The available time for authentication \( T_{\text{auth}} \) is determined by the time it takes a vehicle to travel the given distance $d$, given in meters (before overcoming the camera), at the given speed $v$, meters/seconds: $T_{\text{auth}} = \frac{d}{v}$
\item The total latency \( T_{\text{latency}} \) for the authentication process is the sum of the time $t_f $ to transmit  \( n \) bits and the computation time $t_c$, both expressed in seconds: $T_{latency} = n \cdot t_f + t_c$
\item To balance latency and the available time for authentication, we set \( T_{\text{latency}} \) equal to \( T_{\text{auth}} \), that can be solved for $n$ as $$n \leq \frac{\frac{d}{v} - t_c}{t_f},$$ where \textit{n} represents the maximum possible number of bits that can be flashed.
\end{itemize}

\subsubsection*{Test bed}
Another crucial parameter is the duration of the flash ($t_f$). According to the target scenario, it is possible to find an optimal $t_f$ balance between performance and reliability. 
Using this duration, we can implement the previously presented \textit{n}=14-bit flashing scheme (7 vehicle flashes total) and maintain the sequence close to the flash of one second ($\frac{n}{2} \cdot t_f$).

We consider, for example, a vehicle at a distance of 25 meters from the camera and traveling at a speed of 8.3~m/s, the vehicle has a maximum available time to complete the authentication process ($T_{\text{auth}}$) of approximately 3 seconds using this scheme. At a higher speed of 16.6~m/s, the same distance results in an available $T_{\text{auth}}$ of $ 1.5$ seconds. Despite these variations, we identify in our configuration that the optimal flash duration is $t_f = 0.15$ seconds, which performs reliably in both cases.

\subsection{The Optical Camera
Communication channel}\label{sec:occ}
Our scheme uses the \ac{OCC} model as the \ac{LOS} channel, similar to the one described in \cite{visual1}, where, however, it was used in a different visual authentication context.
In \ac{OCC}, we use \ac{OOK} modulation, as it requires a
single narrowband frequency.
Our design, apart from introducing a novel authentication scheme, adopts distinct trade-offs in timing and security constraints, as illustrated below.

\subsubsection*{Channel Model Parameters}
In \cite{visual1}, the \ac{OCC} channel is modeled with a camera exposure time (\( T_e \)), inversely proportional to the frame rate (\( \text{FPS} \)), where \( T_e = \frac{1}{\text{FPS}} \). 
The pulse width of a transmitted symbol (\( PW_s \)) satisfies \( PW_s < T_e \), ensuring efficient transmission without overlap. 
A key design factor is the duty cycle (\( DC \)), where \( DC_{\min} = \frac{PW_s}{T_e} \), representing the minimum active light duration. Moreover, a guard width (\( PW_g \)) is incorporated to prevent inter-symbol interference, ensuring \( PW_s + PW_g < T_e \). 
In contrast, our approach deviates from \cite{visual1} by prioritizing practical implementation with standard vehicle headlights. 
In our work, we focus on minimizing the flash duration (\( t_f \)), to align with the operational capabilities of the headlight. 
Our design choices emphasize practicality and adaptability, resulting in a cost-effective, real-world solution that can meet both timing and security requirements while integrating authentication into existing vehicle infrastructure.

\subsubsection*{Synchronization and Guard Bands}
The synchronization strategy in \cite{visual1} is based on precise timing to prevent misalignment of the symbols. Transmission times (\( T_i \)) are calculated as \( T_i = st_{i+1} - (PW_s + PW_g) \), where $st_{i+1}$ is the start time of the frame at time \textit{\small{i+1}}, ensuring symbols are transmitted just before the next frame starts. Although this minimizes attack windows, it imposes strict timing constraints.
In contrast, our proposal addresses synchronization differently by structuring the preamble (11-00) and interrupts (00) directly into the flashing frame, avoiding dependence on strict frame synchronization. 
For example, embedding multiple interrupts within the payload ensures symbol distinction even under challenging conditions.
\subsubsection*{Capturing and Processing the Optical Challenge Response}
The flashing can be captured by a camera mounted on an \ac{RSU}, which processes the optical signal to extract the security frame. To detect and interpret the response to the challenge, we employed an \ac{AI}-based approach, leveraging a SlowFast network \cite{slowFast} for robust feature extraction and classification.

\section{Security Analysis} \label{sec:secanalysis}

We evaluate our scheme using the attack model defined in \cref{sec:reference_scenarios}.

\begin{itemize} \item {\it Remote Impersonation Attacks (S, T):}
Attackers may inject falsified messages to impersonate vehicles, exploiting weaknesses in the authentication scheme.
Our dual-channel approach mitigates this threat by combining: \begin{itemize} \item \ac{PKI}-based \ac{NLOS} communication, encrypted with TLS (per IEEE 1609.2, assuming a trusted RA), for cryptographic security. \item \ac{LOS} visual verification, requiring physical presence. \end{itemize}

Authentication requires a valid visual response to a challenge, ensuring that only physically present vehicles are authenticated. Remote attackers who lack physical presence and the correct response cannot succeed. The integrity of the response depends on maintaining a \ac{LOS} link between the vehicle and the camera. Our short execution times simplify this, making the system both practical and efficient.

\item {\it Proximity-Based Attacks (S, E, T):}
Such attacks are critical in restricted-access environments. Our dual-channel scheme counters spoofing and tampering by requiring both credential validation and physical presence before issuing authentication tokens.
This ensures that only legitimate entities gain access to services. The integration of cryptographic and physical verification limits the risk of privilege escalation within the network.

\item {\it Signal Preemption Exploitation (S, T, D):}
In green wave systems, attackers can manipulate signal priorities to disrupt traffic flow.
Although our approach cannot fully eliminate \ac{DoS} risks, it minimizes their effect through lightweight \ac{LOS} authentication, enabling fast processing under load.
Attempts to disrupt the camera (e.g., obstruction or light interference) affect only single-vehicle authentication. As no token is issued without successful verification, such attacks remain isolated and do not affect the broader system. \end{itemize}

\section{Implementation setup} \label{sec:setup}

\begin{figure*}[t!]
    \centering
    \includegraphics[width=\linewidth]{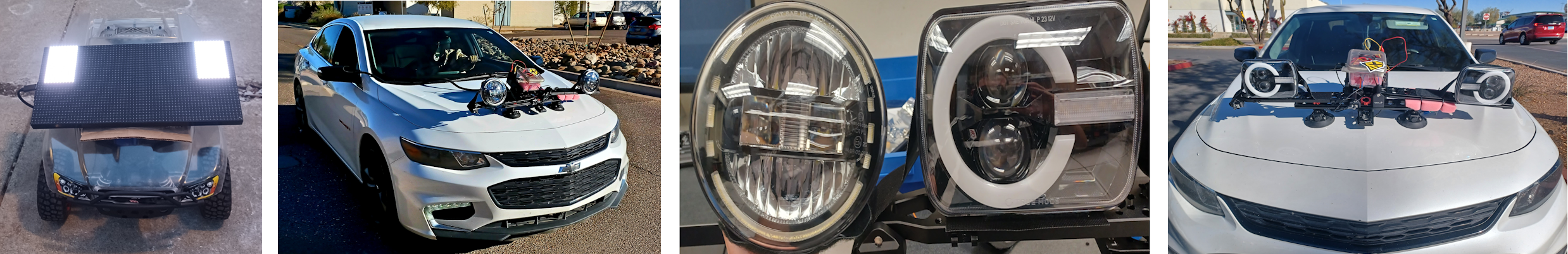}
    \caption{Implementation of our vehicle setup with RC-car and real-car, showcasing different headlight configurations.}
    \label{fig:merged}
\end{figure*}

Our implementation consists of two testbeds: the first developed at \ac{MIT} with an RC-car and the second developed at \ac{ASU} with a real-car. For each testbed, we describe the dataset creation process and the architecture of the deep learning classification model used to evaluate our MFA scheme.

\subsection{Testbeds setup}

\begin{figure}[t!]
    \centering
\includegraphics[width=0.436\textwidth]{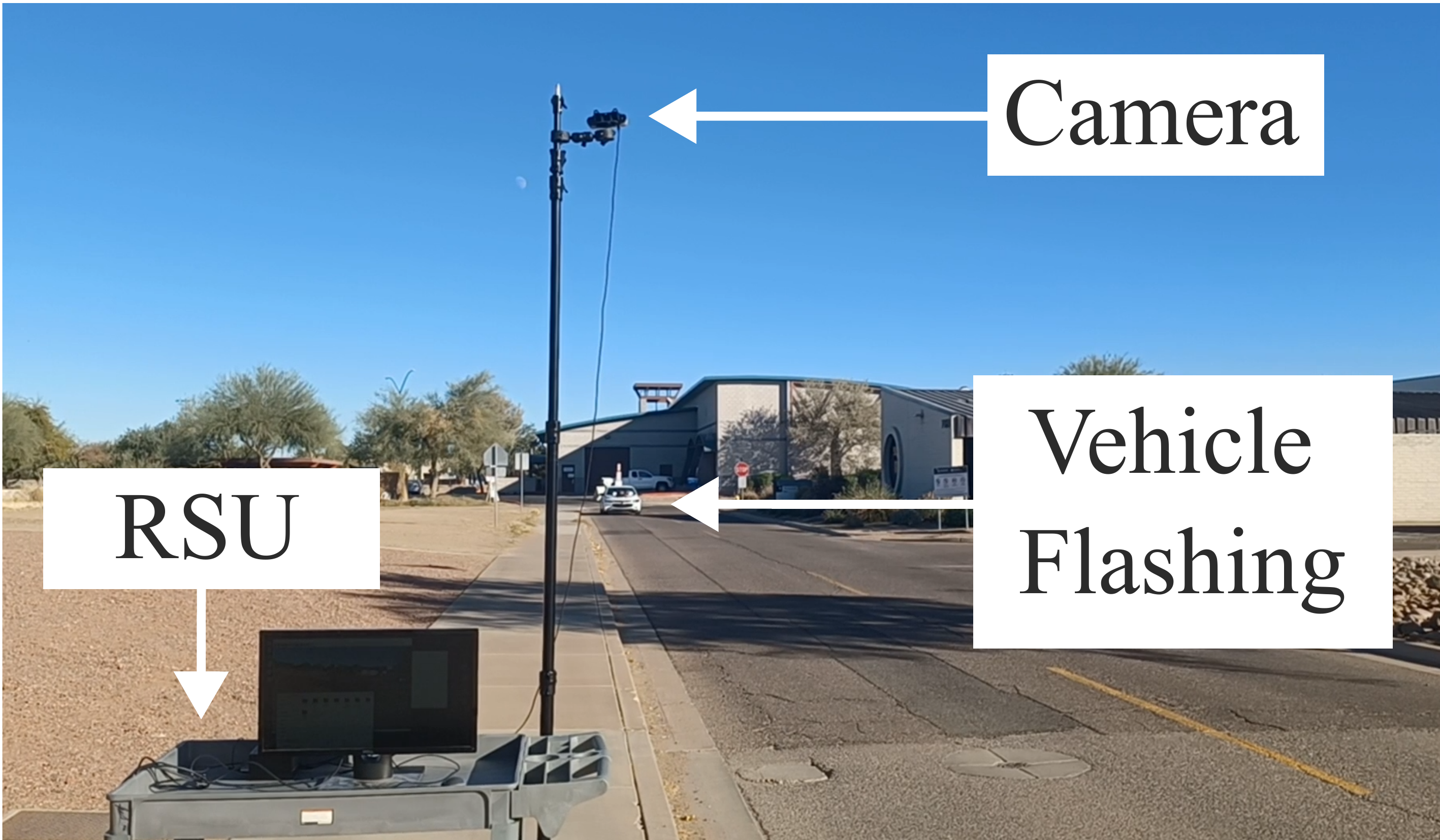}
    \caption{The Phase 2 road set up: \ac{RSU} simulation with camera and edge computer.}
    \label{fig:cameraSetup}
\end{figure}

In both testbed setups, we simulated an \ac{RSU} using an NVIDIA Jetson AGX Orin Developer Kit (P3730) paired with an Intel RealSense Depth Camera D455 (\cref{fig:cameraSetup}). The system was positioned at an elevated height between 1 (RC-car) and 3.5 (real car) meters above ground level. 

\subsubsection{RC-car testbed}  

This setup consists of an RC-car (Traxxas 4W model 58014-4) equipped with an Adafruit M4 Circuit Python-Powered Internet RGB Matrix Display, containing the flashing code, as shown in \cref{fig:merged}. The display consisted of a 64×32 RGB \ac{LED} matrix panel, positioned at the front of the vehicle to simulate dual headlights.

\subsubsection{Real-car testbed} it is 
%was conducted at \ac{ASU} in
a real-world test environment, including public roads and parking lots. It consists of a Chevrolet Malibu equipped with additional headlights to flash the security frame (\cref{fig:merged}). 
We tested two types of headlights with different shapes:
\begin{itemize}
\item Xprite 7" LED Round Headlights and 
\item TRUE MODS 5×7 7×6 Inch H6054 Black LED headlights H4 sealed beam,
\end{itemize}
To control the flashing sequences, we designed an electronic circuit using an Arduino Nano microcontroller. This acted as an intermediary between the vehicle's infotainment system (emulated by a computer) and the headlights. The circuit incorporated MOSFET and other components to regulate power and signal transmission. The Arduino Nano was chosen for its compact form factor, while an external 12V lead-acid battery powered the headlights. The system communicated via UART and received commands through a serial monitor to trigger the appropriate flashing sequences.

\subsection{Datasets Creation} \label{sec:datasets}
\begin{figure*}[t!]
  \centering

  \begin{subfigure}[t]{0.24\textwidth}
    \includegraphics[width=\textwidth]{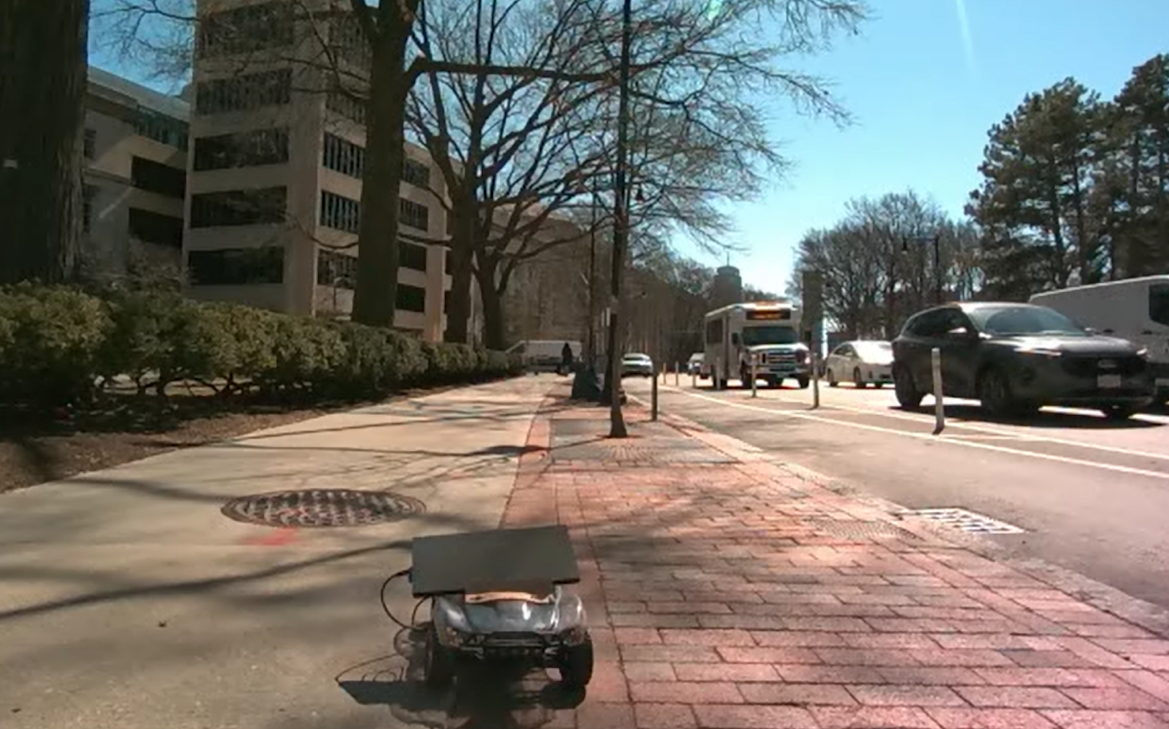} 
    \caption{Urban Road: day/sunny.}
    \label{fig:day_mass2}
  \end{subfigure}
  \hfill
  \begin{subfigure}[t]{0.24\textwidth}
    \includegraphics[width=\textwidth]{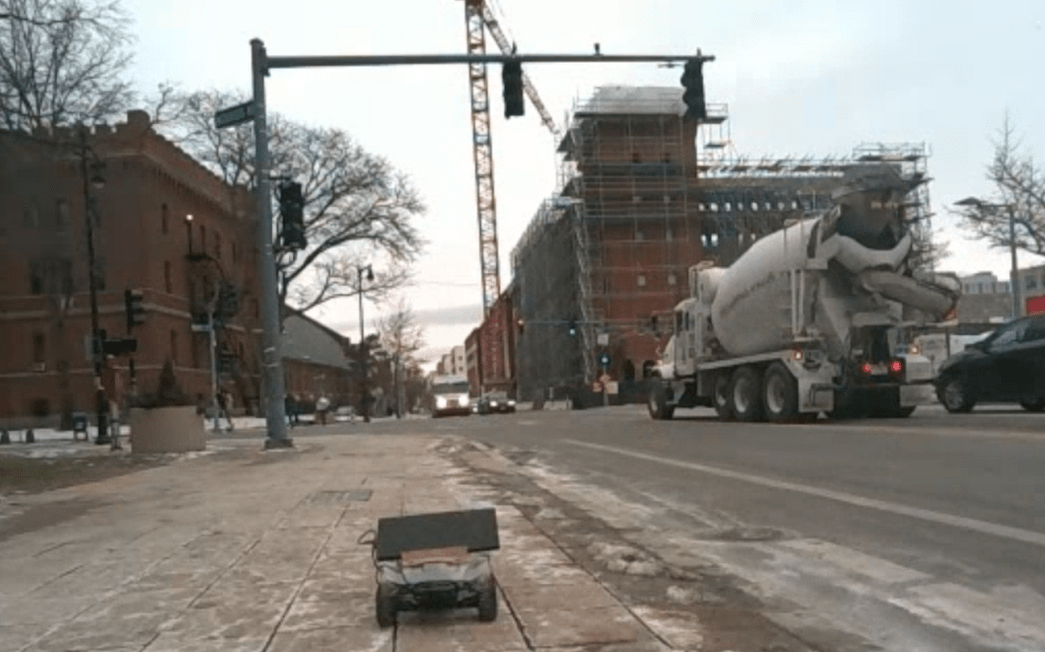}
    \caption{Urban Intersection: day/cloudy.}
    \label{fig:day_vassar2}
  \end{subfigure}
  \hfill
  \begin{subfigure}[t]{0.24\textwidth}
    \includegraphics[width=\textwidth]{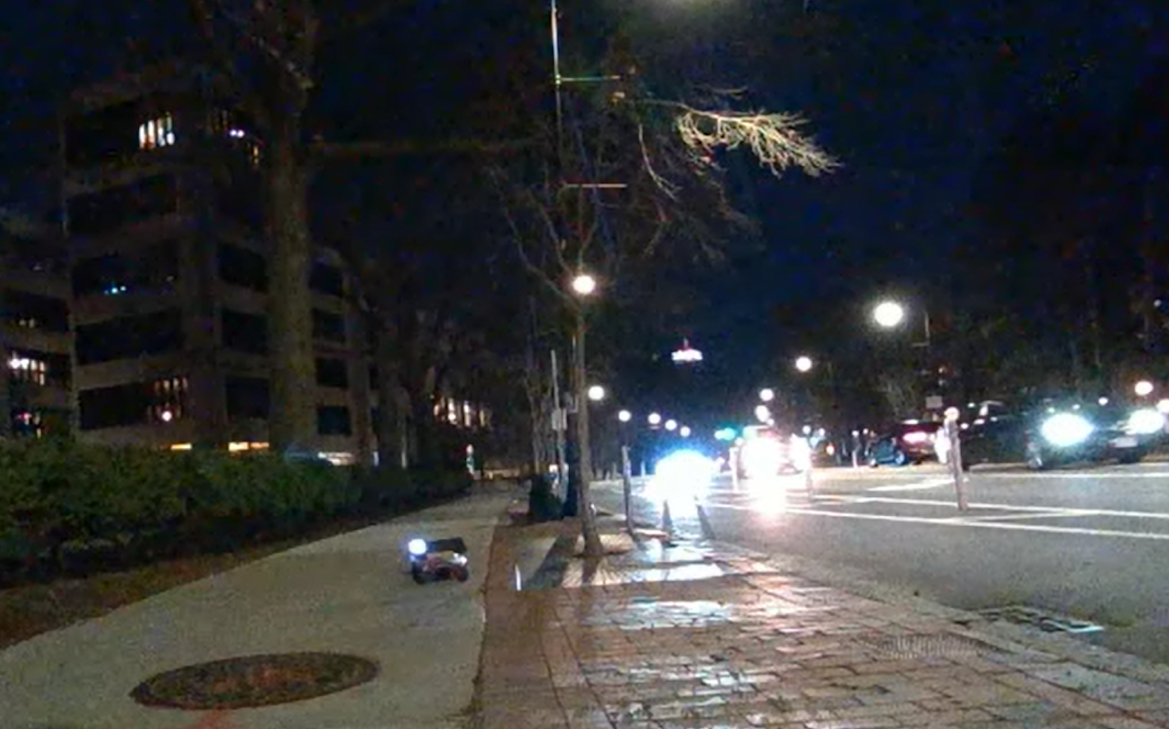}
    \caption{Urban Road: night.}
    \label{fig:night_mass}
  \end{subfigure}
  \hfill
  \begin{subfigure}[t]{0.24\textwidth}
    \includegraphics[width=\textwidth]{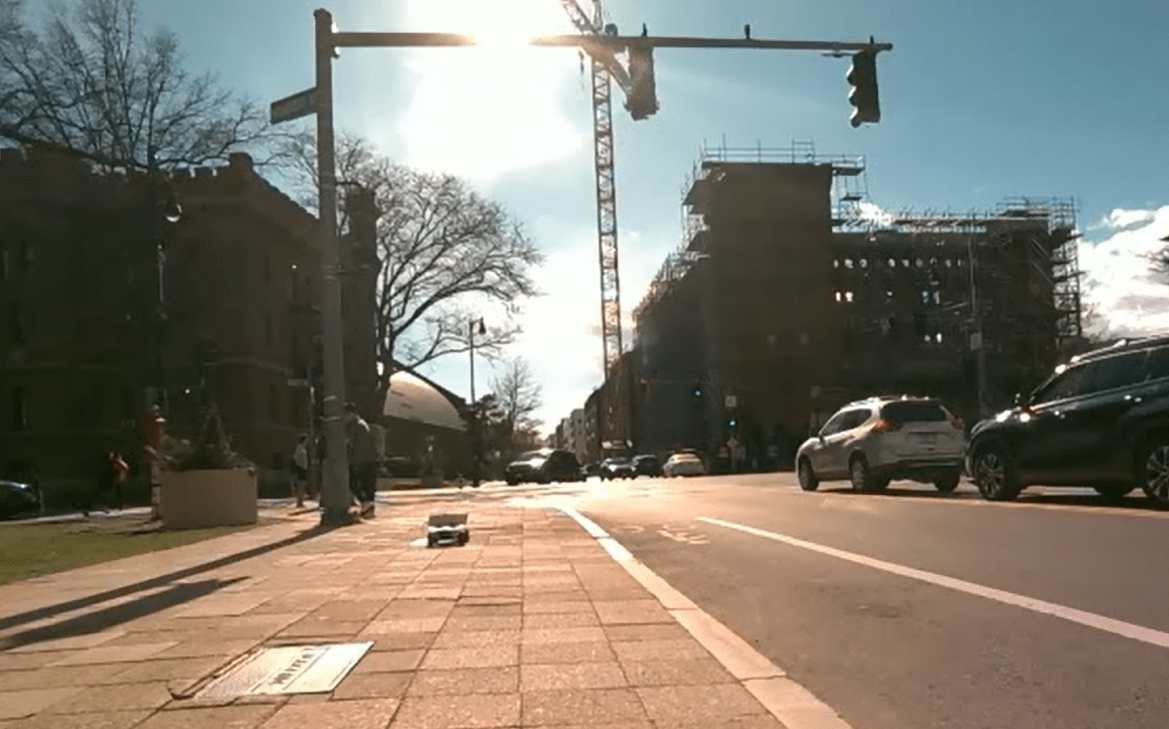}
    \caption{Urban Intersection: day/sunny.}
    \label{fig:day_vassar}
  \end{subfigure}

  \bigskip

  \begin{subfigure}[t]{0.24\textwidth}
    \includegraphics[width=\textwidth]{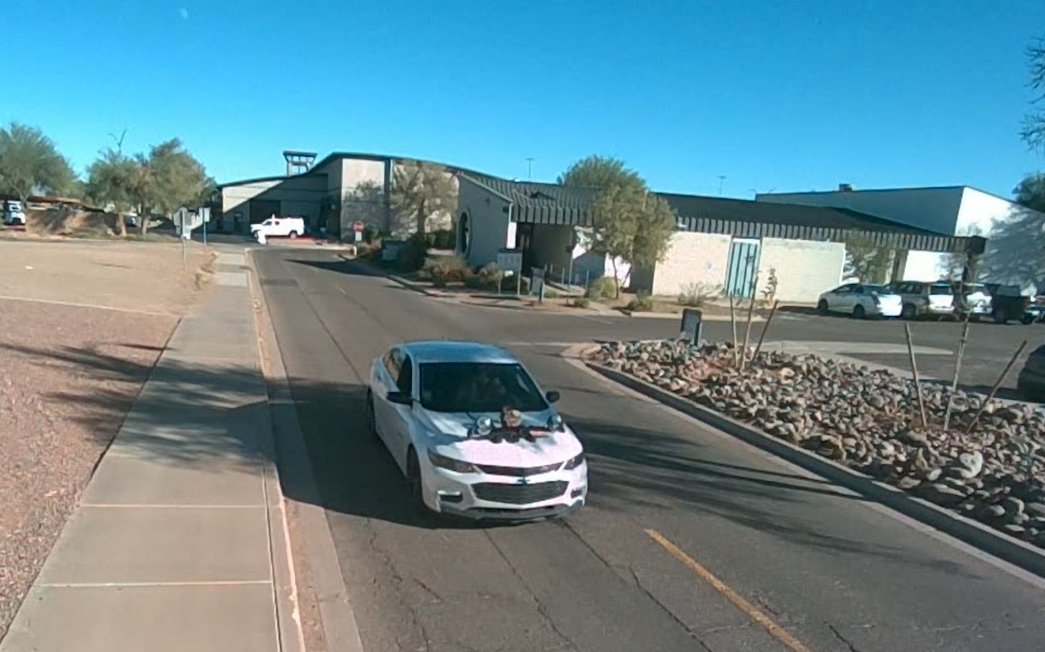}
    \caption{Public Road: day/sunny.}
    \label{fig:day_ASU}
  \end{subfigure}
  \hfill
  \begin{subfigure}[t]{0.24\textwidth}
    \includegraphics[width=\textwidth]{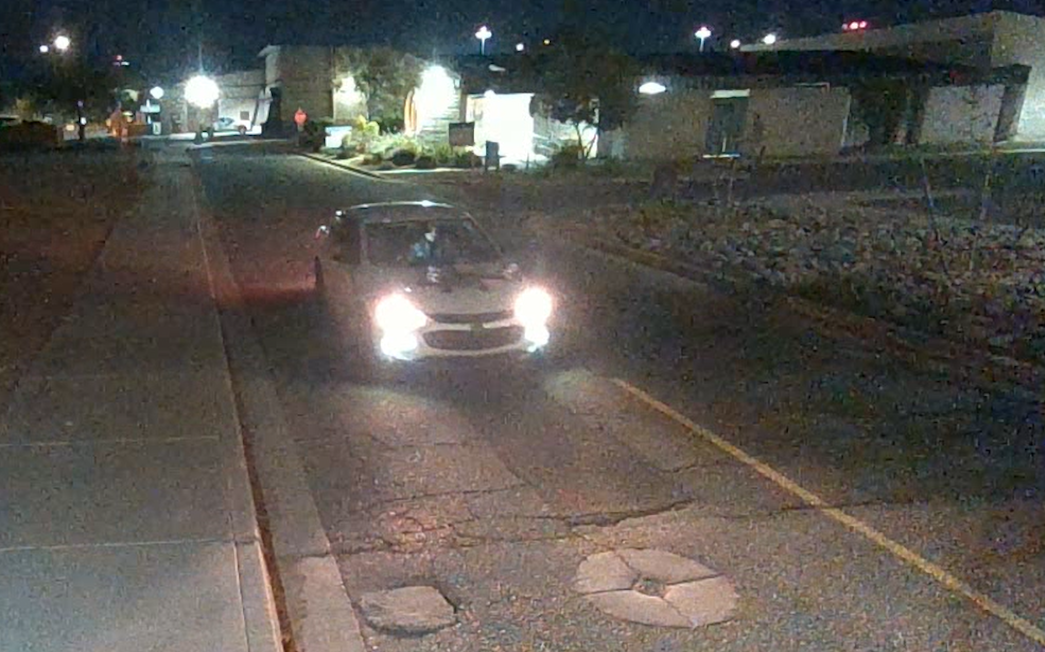} 
    \caption{Public Road: night.}
    \label{fig:night_ASU}
  \end{subfigure}
  \hfill
  \begin{subfigure}[t]{0.24\textwidth}
    \includegraphics[width=\textwidth]{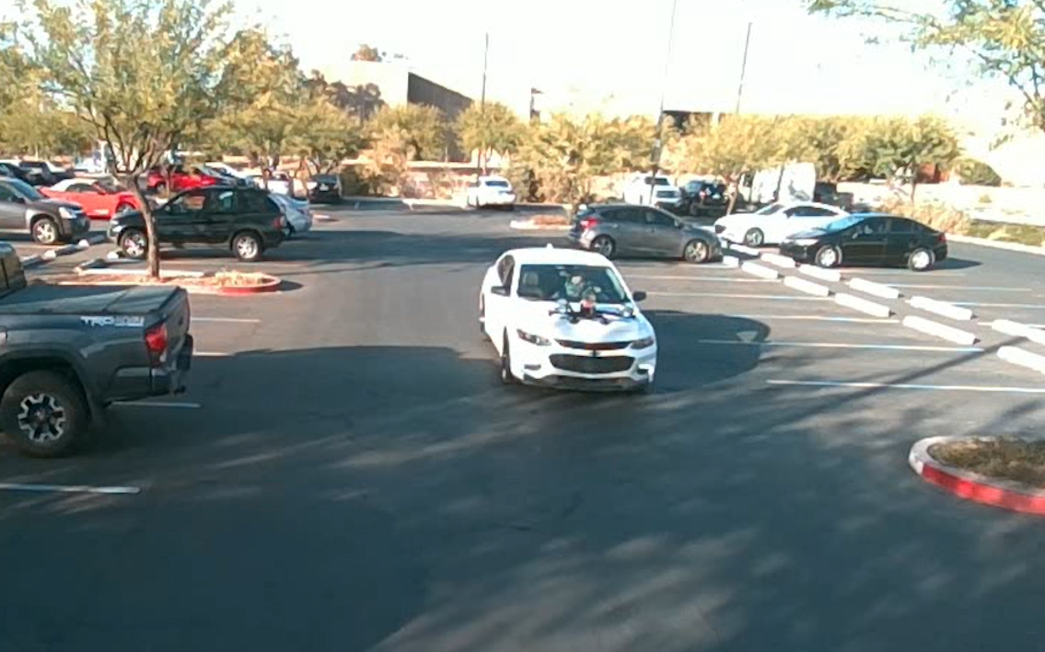} 
    \caption{Parking Lot: day/sunny.}
    \label{fig:day_ASU2}
  \end{subfigure}
  \hfill
  \begin{subfigure}[t]{0.24\textwidth}
    \includegraphics[width=\textwidth]{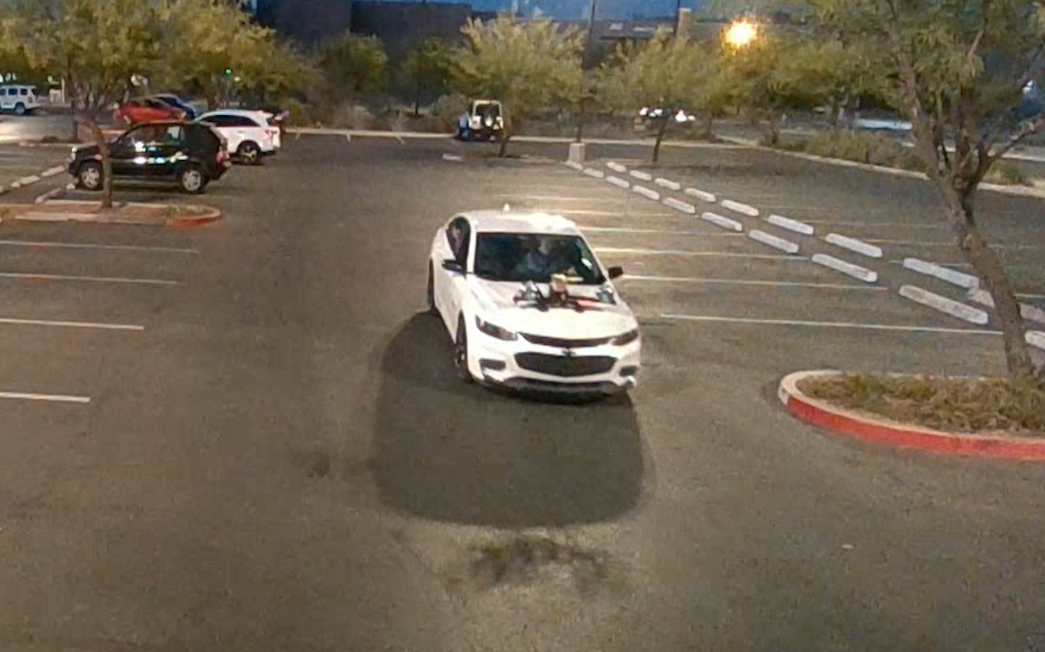}
    \caption{Parking Lot: sunset/night.}
    \label{fig:night_ASU2}
  \end{subfigure}

  \caption{Overview of our primary day/night dataset scenarios. The first row corresponds to Phase 1, and the second to Phase 2.}
  \label{fig:scenarios}
\end{figure*}

To generate datasets for our model, we recorded videos of the authentication process on each testbed. During testing, the vehicle remained in motion in both testbeds. 
In addition, to account for the variability in the real-world, as shown in \cref{fig:scenarios}, we captured videos during the day and night under different lighting conditions.

\subsubsection{RC-car testbed} 
\begin{figure}[t!]
    \centering
    \includegraphics[width=0.85\linewidth]{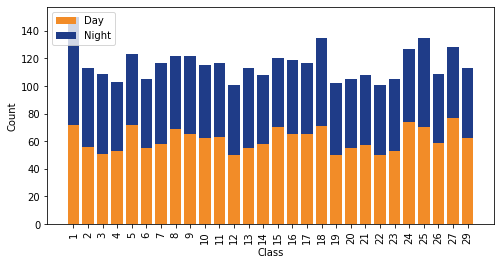}
    \caption{Class distribution for the RC-car dataset.}
    \label{fig:class_dist}
\end{figure}

Using the RC-car, we conducted tests on two public urban roads and one intersection. To faithfully replicate real-world conditions, the \ac{RSU} and the vehicle were placed in a safe location, close to a heavily crowded public road.
The camera recorded videos up to 7 seconds long, capturing the vehicle in motion as it responded to the \ac{RSU} challenge using its headlights. The distance between the vehicle and the camera ranged from 0 to 20 meters.
We collected a dataset of 3,242 videos, including 1,717 recorded during the day and 1,525 recorded at night. The dataset covers 28 classes, derived from the deep learning model classification process, including 27 security sequences and one all-zero class. These classes are numbered from 1 to 27, with the all zero class assigned to number 29. Each class contains between 100 and 150 videos, with at least 50 daytime and 50 nighttime videos per class. \cref{fig:class_dist} illustrates the class distribution of the dataset.

\subsubsection{Real-car testbed} 
\begin{figure}[t!]
    \centering
    \includegraphics[width=0.85\linewidth]{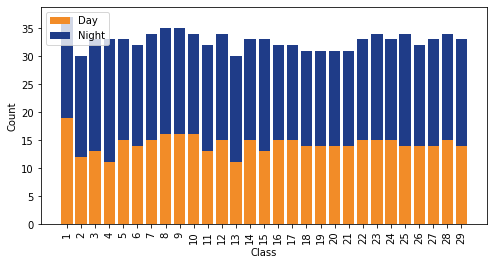}
    \caption{Class distribution for the real-car dataset.}
    \label{fig:class_dist2}
\end{figure}

We conducted realistic and impactful tests by recording videos with a real vehicle moving on a public road and in a parking lot. Compared to the
RC-car testbed, in this case, the distance between the vehicle and the camera ranged from 0 to 50 meters.
In this more realistic test environment, 
the camera recorded videos of a maximum length of 4 seconds, reducing the video dimension and focusing only on the execution of the security frame. We collected a dataset of 975 videos, including 420 recorded during the day and 555 recorded at night. The dataset covers 29 classes, one more than the RC-car dataset, numbered from 1 to 29. Specifically, we introduced class number 28 to represent random flashing patterns that do not conform to predefined security frames, emulating potential real-world anomalies or unexpected vehicle behavior. Each class contains between 30 and 37 videos, with a distribution that varies between daytime and nighttime recordings. \cref{fig:class_dist2} illustrates the class distribution of the dataset. 

\subsection{Headlight Flashing Classification Model Architecture} \label{sec:ml}

To classify the flashing sequences of the headlights in each of the two testbeds we described above, we employed a deep learning approach based on a SlowFast \ac{CNN} architecture \cite{slowFast}.
The model was chosen for its efficiency in processing spatio-temporal features in videos, making it well-suited for recognizing structured patterns in flashing sequences
The SlowFast network consists of two parallel pathways: a slow pathway that captures spatial semantics at a lower frame rate, and a fast pathway that operates at a higher frame rate to detect rapid motion features. This dual-pathway design enables the model to effectively recognize both short-term and long-term temporal dependencies in the headlight flashing sequences. In \cref{tab:slowfast_params}, we report the most relevant parameter we used to design our model.

\begin{table}[ht]
    \centering
    \renewcommand{\arraystretch}{1.1}
    \caption{Summary of SlowFast Model Parameters.}
    \label{tab:slowfast_params}
    \begin{tabular}{|p{4cm}|p{3.9cm}|} 
        \hline
        \textbf{Parameter} & \textbf{Value} \\
        \hline
        \textbf{Model Architecture} & SlowFast R50 \\
        \hline
        \textbf{Pretrained Dataset} & Kinetics-400 \\
        \hline
        \textbf{Backbone} & Dual ResNet-50 Streams \\
        \hline
        \textbf{Input Frames} & 32 \\
        \hline
        \textbf{Normalization} & [0,1] Range \\
        \hline
        \textbf{Slow Pathway Sampling Factor ($\alpha$)} & 4 \\
        \hline
        \textbf{Final Layer} & Fully Connected (Modified) \\
        \hline
        \textbf{Dropout Rate} & 0.2 \\
        \hline
        \textbf{Training/Validation Split} & 80\% / 20\% \\
        \hline
        \textbf{Optimizer} & Adam \\
        \hline
        \textbf{Initial Learning Rate} & $1 \times 10^{-4}$ \\
        \hline
        \textbf{Learning Rate Scheduler} & Warm-up (2 epochs) + Cosine Annealing \\
        \hline
        \textbf{Loss Function} & Categorical Cross-Entropy \\
        \hline
        \textbf{Mixed-Precision Training} & Yes (Gradient Scaling) \\
        \hline
        \textbf{Number of Epochs} & 32 \\
        \hline
        \textbf{Batch Size} & 4 \\
        \hline
    \end{tabular}
\end{table}

\subsubsection{Workflow} Our implementation used the pre-trained SlowFast R50 model from the PyTorchVideo library \cite{pytorchvideo_slowfast}. The backbone of the model consists of two ResNet-50 streams, which provide a strong feature extraction capability. The model was pretrained on Kinetics-400 \cite{kay2017kineticshumanactionvideo}, a large-scale action recognition dataset, which significantly accelerated the learning process by providing useful low-level features. We modified the final projection layer, replacing it with a fully connected layer that maps the extracted features to the number of flashing classes in our dataset. A dropout layer with a probability of 0.2 was added to prevent overfitting and improve generalization.

\subsubsection{Dataset preparation}
The dataset was split into 80 percent training and 20 percent validation, ensuring that the model learned effectively from diverse sequences while preventing overfitting. No separate test set was used in this phase. Each video was preprocessed to extract 32 frames, evenly sampled along its duration. The frames were normalized to the [0,1] range and converted to tensor format. To construct the SlowFast input pathways, the Fast pathway received the full frame sequence, while the Slow pathway received a subsampled version with an alpha factor of 4, ensuring that it captured long-term temporal dependencies. This strategy allows the model to balance fine-grained motion detection with a broader contextual understanding of the flashing sequences. Unlike conventional object detection approaches, we did not use any bounding boxes to isolate the headlights; instead, the entire video clip was fed into the network without any additional data augmentation.

\subsubsection{Model Generation}
The model was trained with Adam Optimizer with an initial learning rate of $1\times10^{-4}$. A variable learning rate was implemented, where the first two epochs served as a warm-up period, during which the learning rate gradually increased. Afterwards, a cosine annealing scheduler was applied to smoothly decay the learning rate, improving stability and convergence. The loss function was categorical cross-entropy. Mixed-precision training was used using gradient scaling to optimize memory efficiency and computational speed on the GPU. The training spanned 32 epochs, with a batch size of 4 to accommodate GPU memory constraints. The accuracy of the validation was monitored at each epoch, and the model checkpoints were saved to preserve the best-performing weights.

\subsection{Results}\label{sec:MLresults}
{Let us now compare the results we obtained on both the considered testbeds, RC-car and real-car, to evaluate the proposed MFA scheme. In particular, we measure data loss and accuracy on the training dataset.  
Then, in both cases we conducted separate testing rounds per dataset.
\cref{tab:comparisonImplem} provides a comparison of the parameters and results.
\begin{table}[t!]
\centering
\footnotesize
\renewcommand{\arraystretch}{1.15}
\caption{Comparison between Phase 1 and 2.}
\begin{tabular}{|p{3cm}|p{2.12cm}|p{2.3cm}|}
\hline
\textbf{Implementation} & \textbf{Phase 1 (RC-car)} & \textbf{Phase 2 (Real-car)} \\
\hline
\textbf{Vehicle} & RC-car & Chevrolet Malibu \\
\hline
\textbf{Light source} & LED screen & LED Headlights \\
\hline
\textbf{\ac{RSU} camera} & \multicolumn{2}{c|}{Intel RealSense D455} \\
\hline
\textbf{Edge Computer} & \multicolumn{2}{c|}{NVIDIA Jetson AGX Orin} \\
\hline
\textbf{Vehicle-\ac{RSU} distance} & 0-20 m & 0-50 m \\
\hline
\textbf{Single flash duration} & \multicolumn{2}{c|}{0.15 s} \\
\hline
\textbf{Day/night videos} & \multicolumn{2}{c|}{Yes} \\
\hline
\textbf{Settings} & 3 & 2 \\
\hline
\textbf{Classes} & 28 (27 + 0 s) & 29 (27 + 0 s + random flash) \\
\hline
\textbf{Video length} & 2-6 s & 1-3 s \\
\hline
\textbf{Collected videos} & 3242 & 975 \\
\hline
\textbf{ML model} & \multicolumn{2}{c|}{CNN} \\
\hline
\textbf{Best accuracy on test set} & 96.1\% & 97.9\% \\
\hline
\textbf{Average accuracy on test} & 95\% & 96.6\% \\
\hline
\end{tabular}
\label{tab:comparisonImplem}
\end{table}

\begin{itemize}
    \item 
{\bf RC-car testbed} 
It reached the best results at 14 epochs.
The model achieved test accuracies of 95.29\%, 94.47\%, 96.11\%, 94.67\%, and 94.67\%, respectively. The model achieved a recall of 0.95, and F1-score of 0.96 across 28 classes.

\item
{\bf Real-car testbed} 
It reached the best results at 32 epochs.
The model achieved test accuracies of 97.89\%, 95.77\%, 97.89\%, 94.37\%, and 97.18\%, respectively. The model achieved a recall of 0.96, and F1-score of 0.96 across 29 classes.
\end{itemize}

In both testbeds, the model demonstrated consistently high performance with minimal accuracy fluctuations across different test splits. The average accuracy converged to approximately 95.04\% (RC-car) and 96.6\% (real-car), highlighting its reliability in challenging scenarios.
\begin{itemize}
    \item {\bf RC-car}: The model effectively handled sudden light changes, close traffic, varying external lights, and distances up to 20 meters.
    \item {\bf Real-car}: The model showed robustness to day/night transitions and distances up to 50 meters.
\end{itemize}

In our context, a misclassification includes both true negatives (TN), where a vehicle flashes the correct pattern but it is misclassified, and false positives (FP) on invalid attempts, where a vehicle flashes an incorrect pattern and the model correctly detects it. In the real-car implementation, across five test rounds totaling 730 video clips, the model misclassified approximately 3.40\% of attempts. These include both TN cases, where valid flashes were rejected, and FP cases on invalid flashes that were rightly flagged.

\noindent
{\bf Training Dynamics}: \Cref{fig:accuracy_curve_MIT} and \Cref{fig:accuracy_curve_ASU} show how validation accuracy steadily improves over training epochs, stabilizing near 95\%, where it achieves the best accuracy on the
sets. 
\Cref{fig:loss_curve_MIT} and \Cref{fig:loss_curve_ASU} illustrate the corresponding training and validation loss curves, both converging to minimal values, indicating effective learning with limited overfitting. The small gap between training and validation curves further suggests strong generalization across different environments.

\begin{figure}[ht!]
  \centering
\begin{subfigure}[t]{0.24\textwidth}
    \includegraphics[width=\textwidth]{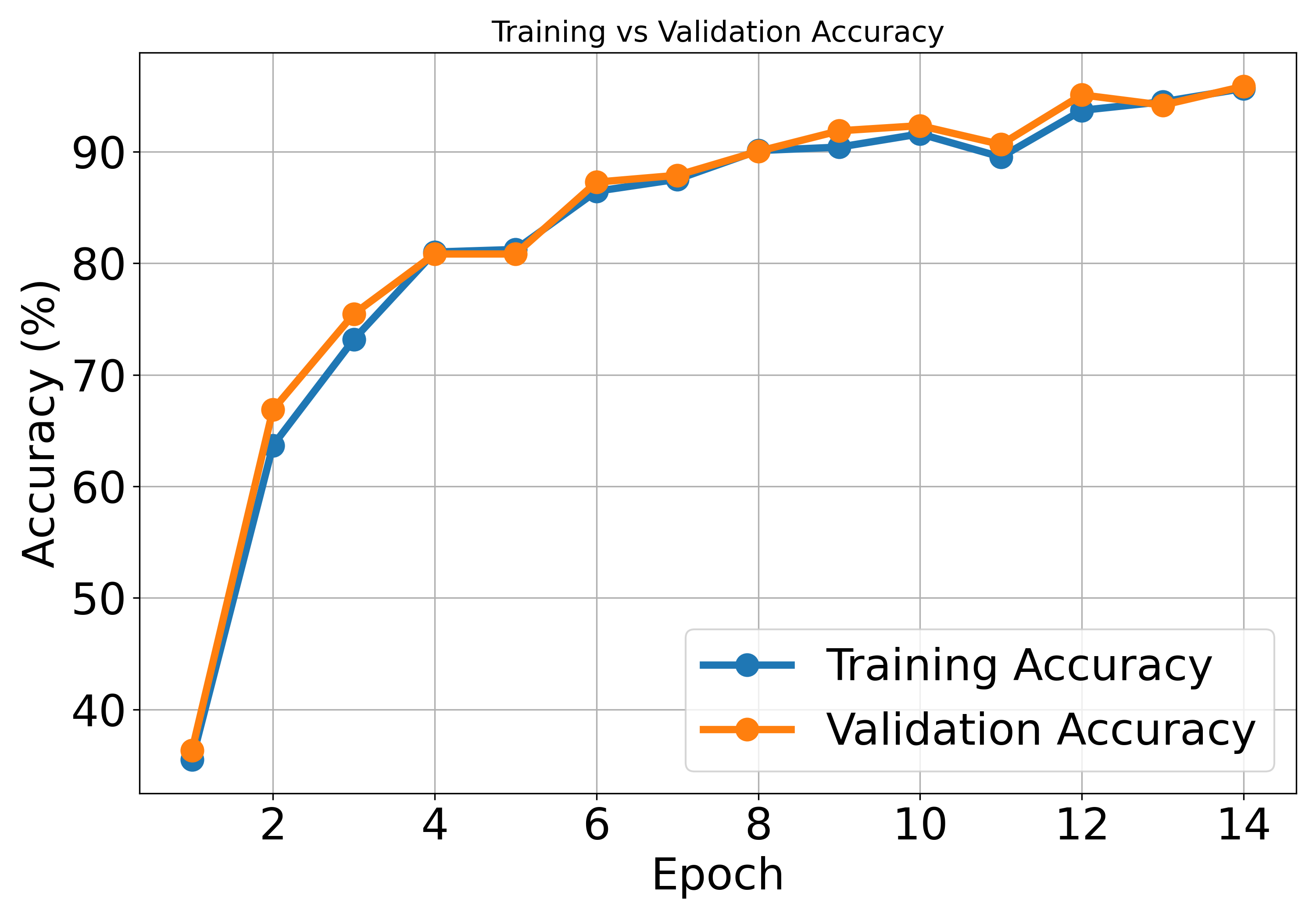} 
    \caption{RC-car dataset training and validation accuracy curves.}
    \label{fig:accuracy_curve_MIT}
  \end{subfigure}
  \hfill
  \begin{subfigure}[t]{0.24\textwidth}
    \includegraphics[width=\textwidth]{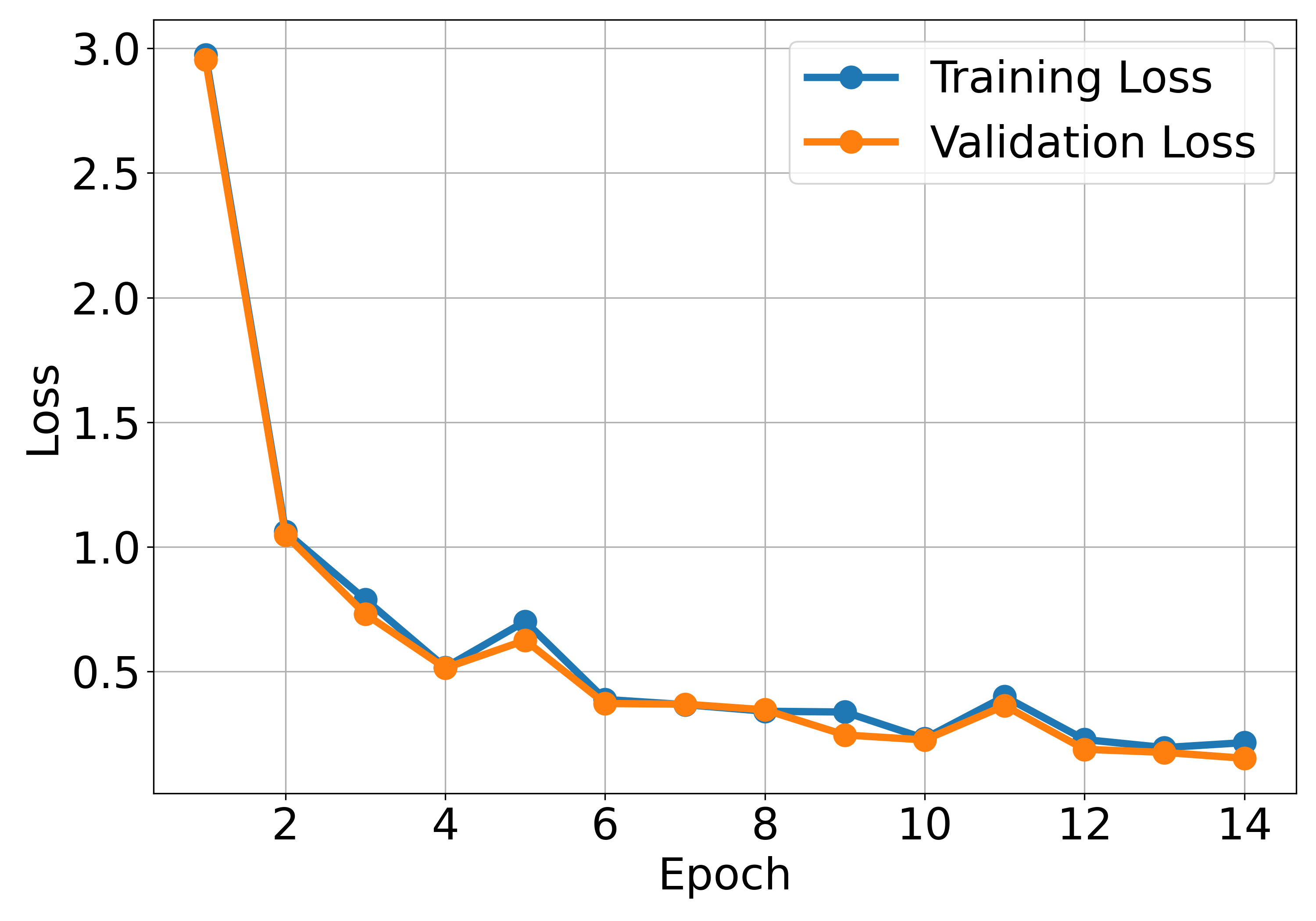}
    \caption{RC-car dataset training and validation loss curves.}
    \label{fig:loss_curve_MIT}
  \end{subfigure}
  \hfill
  \begin{subfigure}[t]{0.24\textwidth}
    \includegraphics[width=\textwidth]{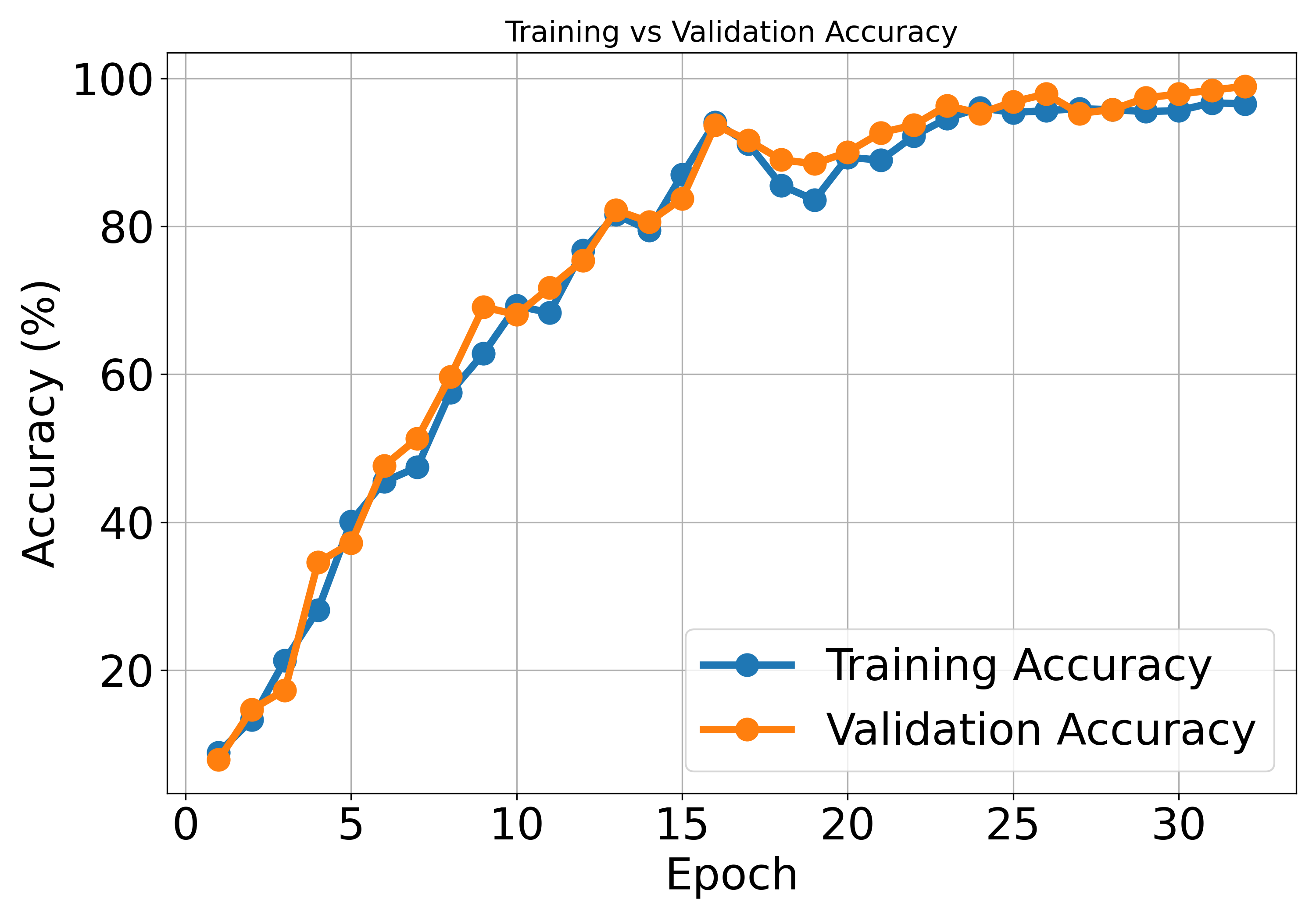}
    \caption{Real-car dataset training and validation accuracy curves.}
    \label{fig:accuracy_curve_ASU}
  \end{subfigure}
  \hfill
  \begin{subfigure}[t]{0.24\textwidth}
    \includegraphics[width=\textwidth]{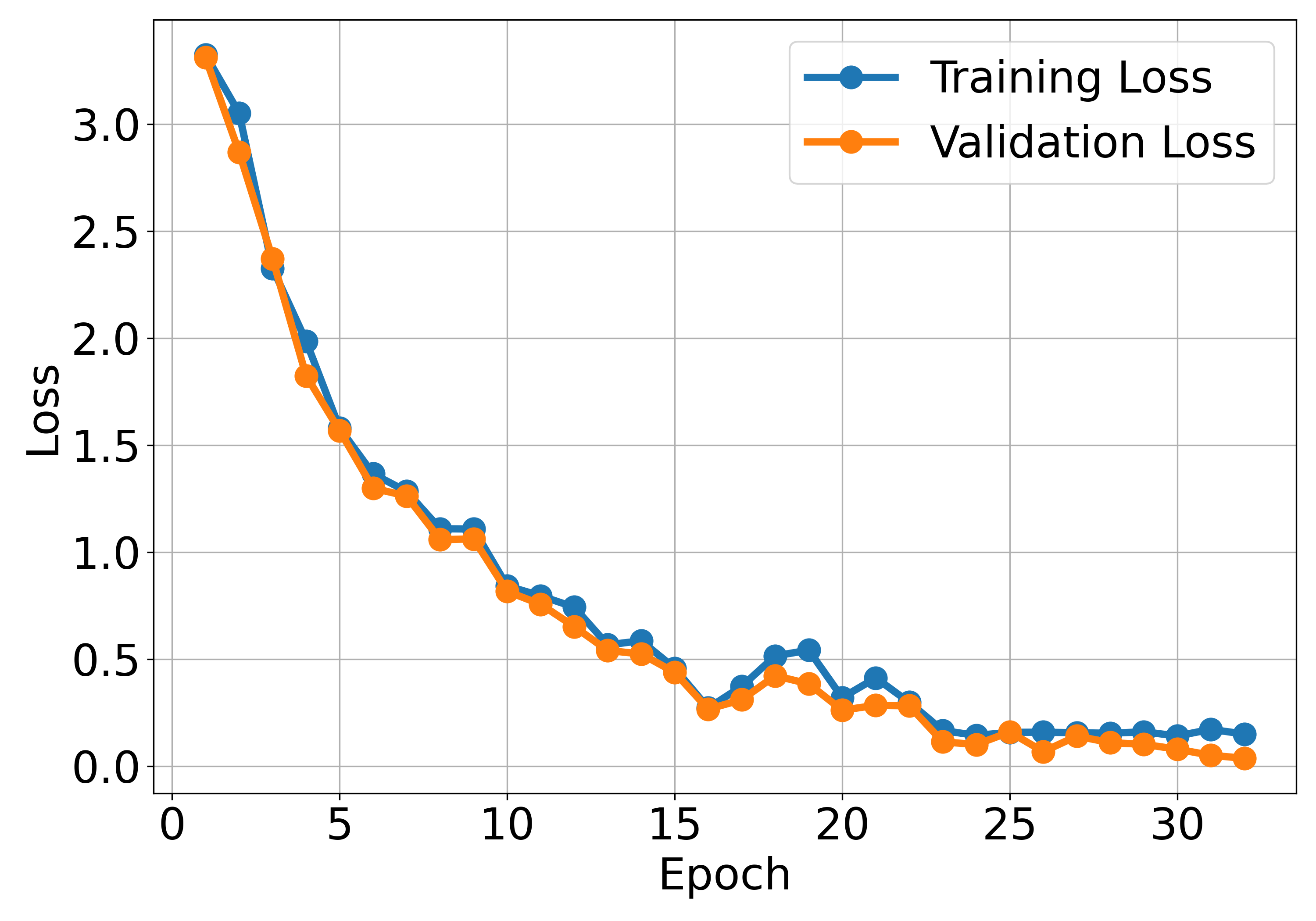} 
    \caption{Real-car dataset training and validation loss curves.}
    \label{fig:loss_curve_ASU}
  \end{subfigure}

  \caption{Loss and accuracy curves of our tests.}
  \label{fig:LossAccuracyCurves}
\end{figure}

\subsection{Models Comparison}\label{sec:MLcomparison}
We compare the performance of three distinct visual-based authentication models used for our same purpose: Dwyer et al.~\cite{benDw}, a 3D CNN initially developed by us, and our final SlowFast CNN. Each model is evaluated based on accuracy and pattern recognition speed.

Each model employs different methods for recognizing authentication patterns, resulting in varying degrees of effectiveness. The first model by Dwyer et al.~\cite{benDw} introduces an \ac{MFA} scheme conceptually similar to ours but based on QR codes displayed by vehicles. As shown in the upper part of \cref{fig:MLcomparison}, this model processes raw video input, utilizes YOLO Object detection to detect the vehicle, and YOLOv8 algorithm to read QR codes displayed by vehicles, and subsequently generates an output in the form of a matrix array representing the recognized authentication pattern. As stated in the article, this approach achieved variable accuracy, ranging from 42\% to 100\% according to the dimension of the QR code and the distance (over 7 m the declared accuracy was 0\%). However, the authors did not explicitly report the pattern recognition times. Therefore, we conducted a test using the same YOLOv8 model on our machine (Intel i9-12900KF 3.19 GHz, 128 GB RAM) and found that the complete pattern recognition process takes between 75 ms and 100 ms, or more in some cases.

In our initial experiments, we developed a 3-layer 3D CNN, designed to decode security messages encoded within vehicle headlight flashing sequences. As shown in \cref{fig:MLcomparison}, this model leverages spatiotemporal convolutions (3D) to effectively capture both the temporal dynamics and spatial characteristics of flashing patterns. It achieved stable and relatively high accuracy of 85.6\%, with a rapid and consistent recognition speed of approximately 1 millisecond on the RC-car dataset tested with the same previous machine. While this demonstrates potential for real-time authentication, its accuracy significantly deteriorated on the real-car dataset.

To address this limitation, we advanced to the SlowFast CNN architecture, specifically designed to concurrently analyze slow (spatial) and fast (temporal) streams of information. As shown in \cref{fig:MLcomparison}. the pretrained network and the dual-pathway design enables the model to simultaneously interpret detailed visual features and rapidly changing patterns. As a consequence, this model significantly outperformed previous approaches, achieving higher accuracy and reduced latency when tested on a machine with lower computational power compared to the previous one. It also demonstrated strong generalization capabilities across both RC-car and real-car datasets. In summary, as illustrated in \cref{fig:MLcomparison}, our SlowFast CNN model provides an optimal balance between high accuracy and fast, reliable processing, making it well-suited for practical deployment in real-world vehicular authentication scenarios.

\begin{figure}[t!]
    \centering
    \includegraphics[width=0.95\linewidth]{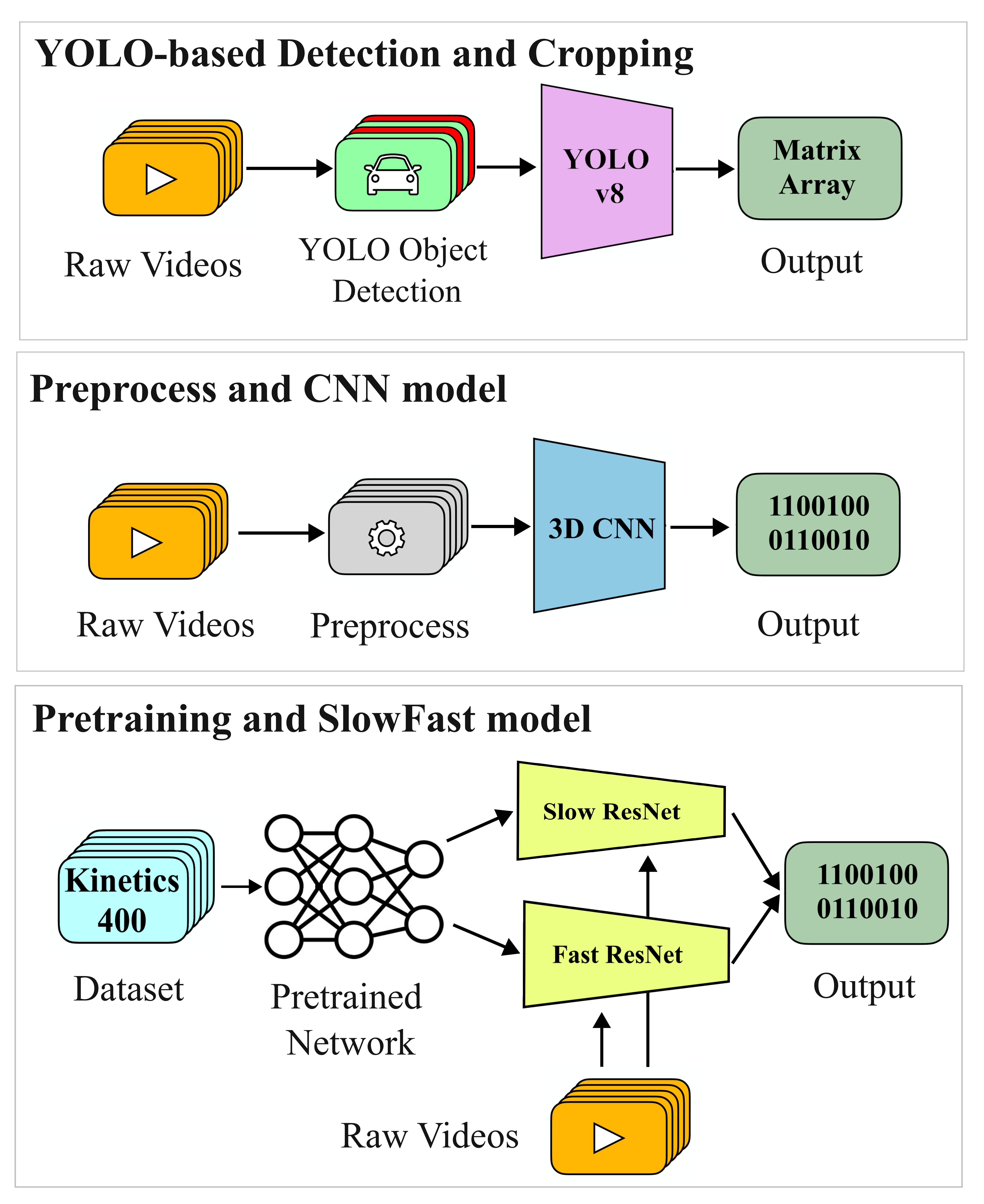}
    \caption{Pipeline comparison among the visual-based authentication models.}
    \label{fig:Pipelinecomparison}
\end{figure}

\begin{figure}[t!]
    \centering
    \includegraphics[width=0.87\linewidth]{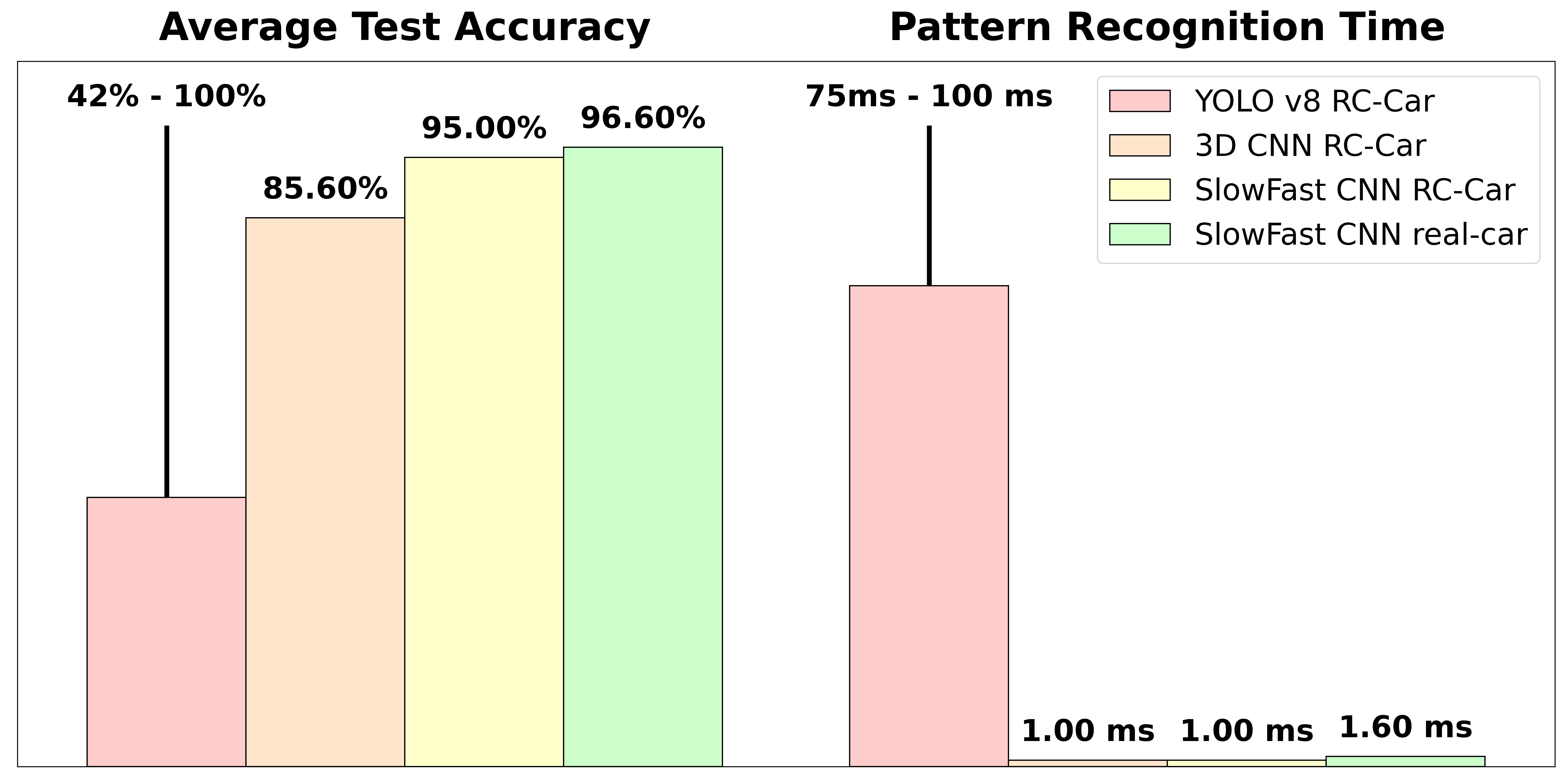}
    \caption{Performance comparison among the visual-based authentication models.}
    \label{fig:MLcomparison}
\end{figure}

\subsection{Model Ablation}\label{sec:MLabaltion}
Based on the network in \cref{fig:slowfast_architecture} and the ablation performed in \cite{slowFast}, where the authors evaluated the impact of the Fast and Slow pathways to highlight their complementary nature, we followed a similar approach for our model. However, we first removed the Slow pipeline, since in our scenario the Fast part captures critical flashing operations, while contextual semantics is less relevant. To perform this modification, we supplied a zero-filled tensor to the Slow branch during training and testing. Additionally, we adapted the data preprocessing and input collation procedures to exclude any handling of Slow pathway data, ensuring that only the Fast stream was actively utilized by the model.

Secondly, we disabled the lateral connections from the Fast to the Slow pipeline during training to evaluate the independence and contribution of each component. In the original architecture, these lateral connections are implemented as convolutional layers that fuse high-temporal-resolution features from the Fast pathway into the Slow pathway at specific depths. To eliminate their influence, we manually zeroed their weights and biases and froze them to prevent further updates during training. This ensured that no information flowed from the Fast stream into the Slow stream, allowing us to assess how well the two branches perform without mutual support or shared features.

\begin{figure}[t!]
    \centering
    \begin{subfigure}[b]{0.95\linewidth}
        \centering
        \includegraphics[width=\linewidth]{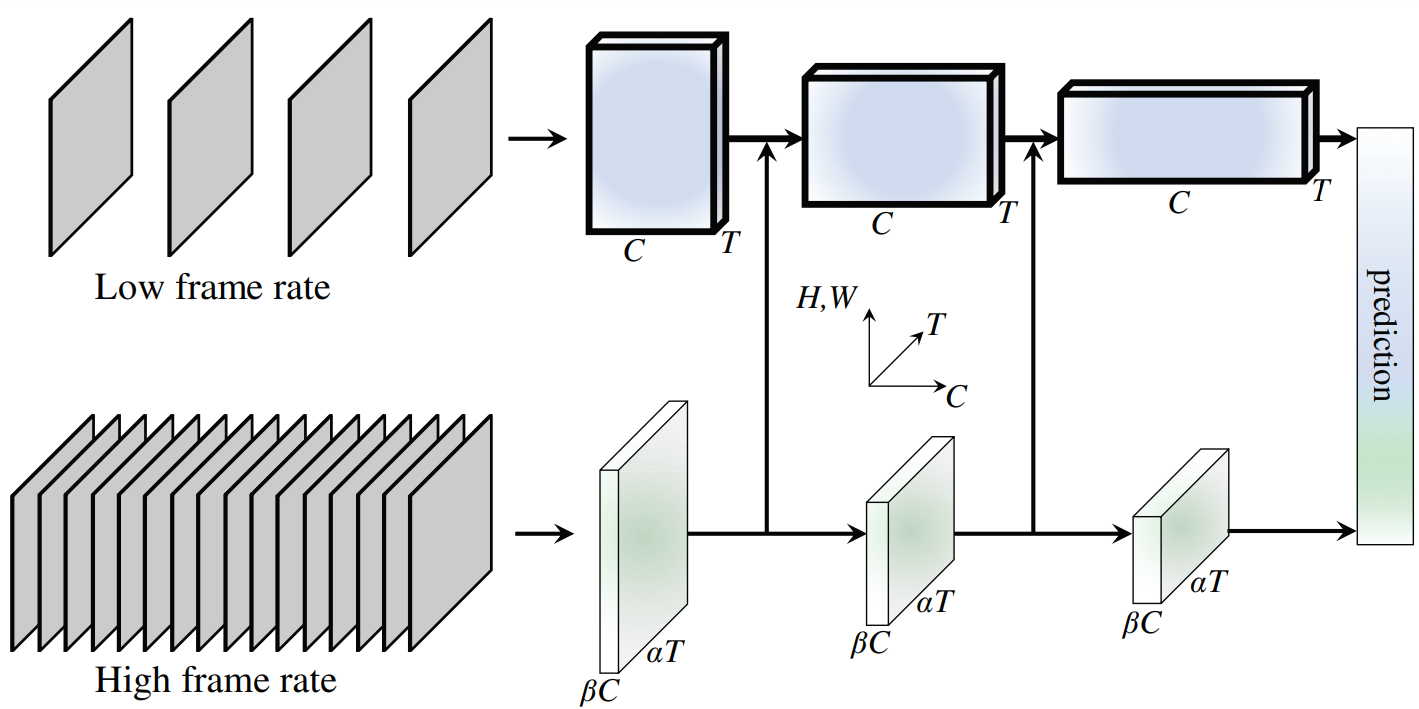}
        \caption{Illustration of the SlowFast network architecture for video recognition \cite{slowFast}.}
        \label{fig:slowfast_architecture}
    \end{subfigure}
    \vskip\baselineskip
    \begin{subfigure}[b]{0.97\linewidth}
        \centering
        \includegraphics[width=\linewidth]{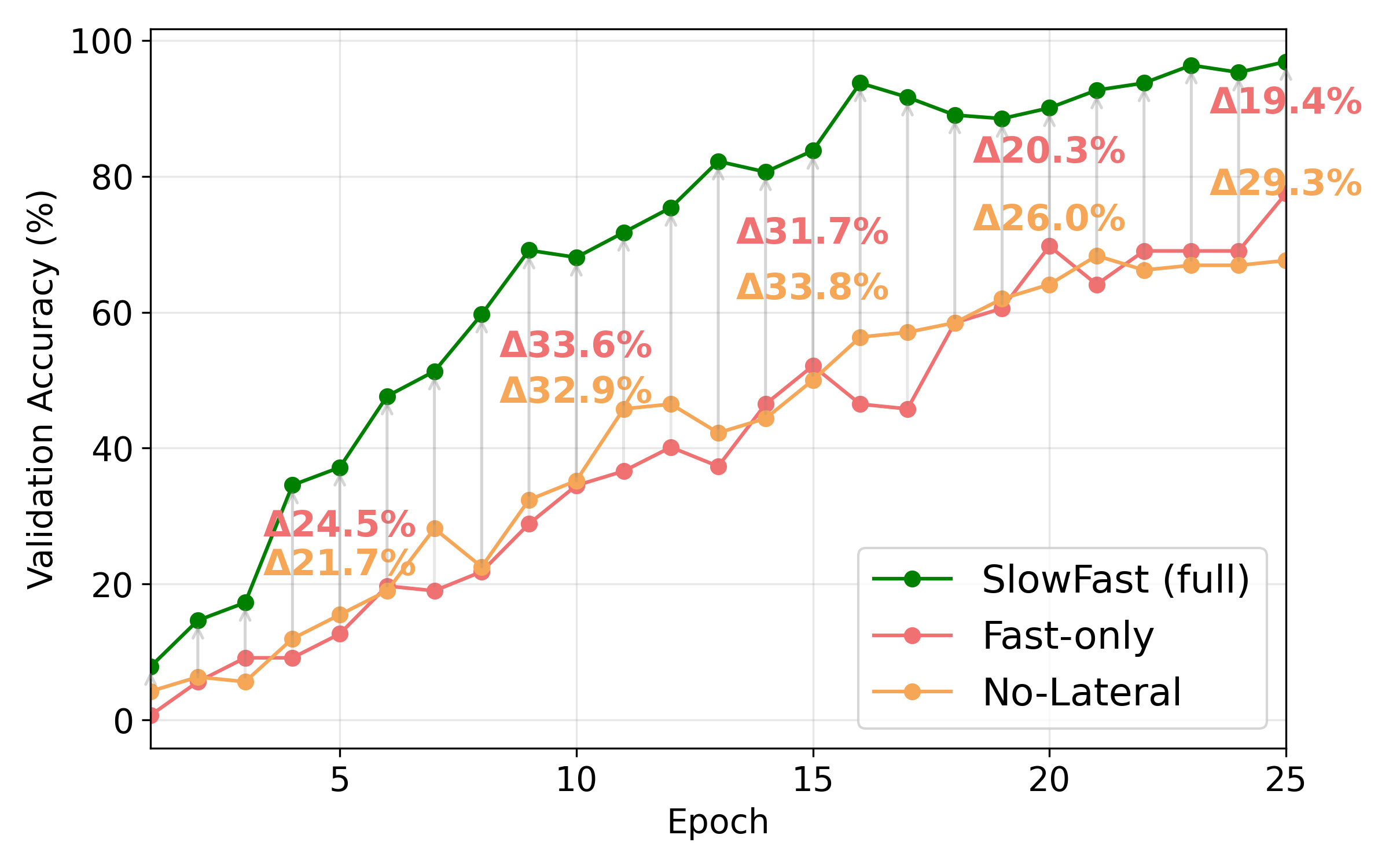}
        \caption{Ablation study comparing validation accuracy for Fast-only, No-Lateral, and full SlowFast networks.}
        \label{fig:slowfast_ablation}
    \end{subfigure}
    \caption{(a) Overview of the SlowFast architecture, showing the slow and fast pathways. (b) Ablation study results demonstrating the importance of both pathways and lateral connections.}
    \label{fig:ablation_plot}
\end{figure}

\begin{comment}
\begin{figure}[htb!]
\centering
\includegraphics[width=0.48\textwidth]{images/val_accuracy_comparison.png}
\caption{Validation accuracy over 25 training epochs for the three configurations. The vertical deltas ($\Delta$) quantify the accuracy difference between the full model and its ablated variants at key epochs.}
\label{fig:ablation_plot}
\end{figure}
\end{comment}

Figure~\ref{fig:ablation_plot} highlights the contribution of each architectural component to the overall performance of the model. The complete SlowFast network achieves nearly 97\% validation accuracy by epoch 25. The Fast-only variant shows a consistent performance gap compared to the full model, with a final delta of $\Delta 19.4\%$ at epoch 25 and then started to degrade. This gap indicates that, while the Fast pathway is essential for capturing rapid temporal features, the contextual information from the Slow branch significantly improves model understanding. The No-Lateral variant underperforms the Fast-only model in most epochs, ending with a gap of $\Delta 29.3\%$ from the full architecture. This suggests that lateral connections, which allow the Fast pathway to enrich the Slow pathway, play a crucial role in learning spatio-temporal relationships and improving feature representation in our application.

\subsection{Discussion}\label{sec:MLdiscussion}

Both testbeds demonstrate the robustness and effectiveness of the proposed headlight flash classification approach. 
Our method reliably operates under diverse conditions, including day and night, varying distances, different headlight shapes (round or rectangular), and even random flashing, highlighting its strong potential for real-world applicability.
By not relying on bounding boxes or strict lighting constraints, the model remains resilient to variations in vehicle types, headlight shapes, and environmental lighting. These results suggest that the proposed approach can be effectively generalized in different contexts, paving the way for a practical and secure \ac{V2I} authentication system that is accurate and efficient.

As in \cite{slowFast}, Fig.~\ref{fig:per_class_ap} illustrates the per-class Average Precision (AP) computed over 29 classes in the real-car testbed. The AP for each class was calculated based on our test set by measuring the ratio of true positives to the sum of true positives and false negatives. The mean Per-class Accuracy (mPA), obtained by averaging the AP values across all classes, ensures that the model's performance is not dominated by more frequent classes but fairly represents all flashing patterns. As shown, the model consistently achieves high AP values in most classes, further supporting the strong recall and F1 scores previously reported. This comprehensive evaluation confirms the robustness and balanced reliability of the model.

\begin{figure*}[t!]
  \centering

  \begin{subfigure}[t]{0.88\textwidth}
    \includegraphics[width=\textwidth]{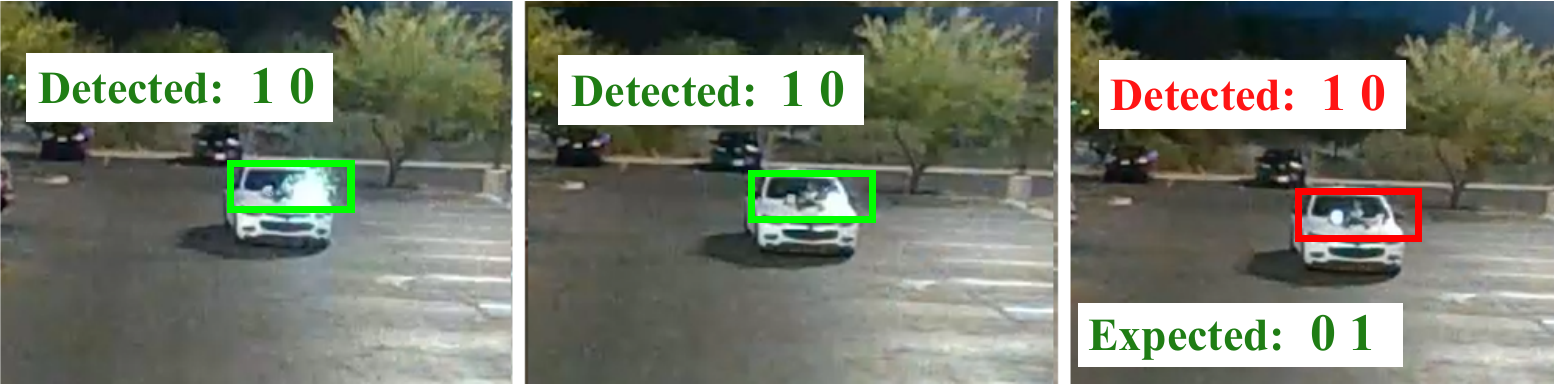} 
    \caption{Class 15 (10-10-01) misclassified as class 14 (10-10-10).}
    \label{fig:miss1}
  \end{subfigure}

   \begin{subfigure}[t]{0.88\textwidth}
    \includegraphics[width=\textwidth]{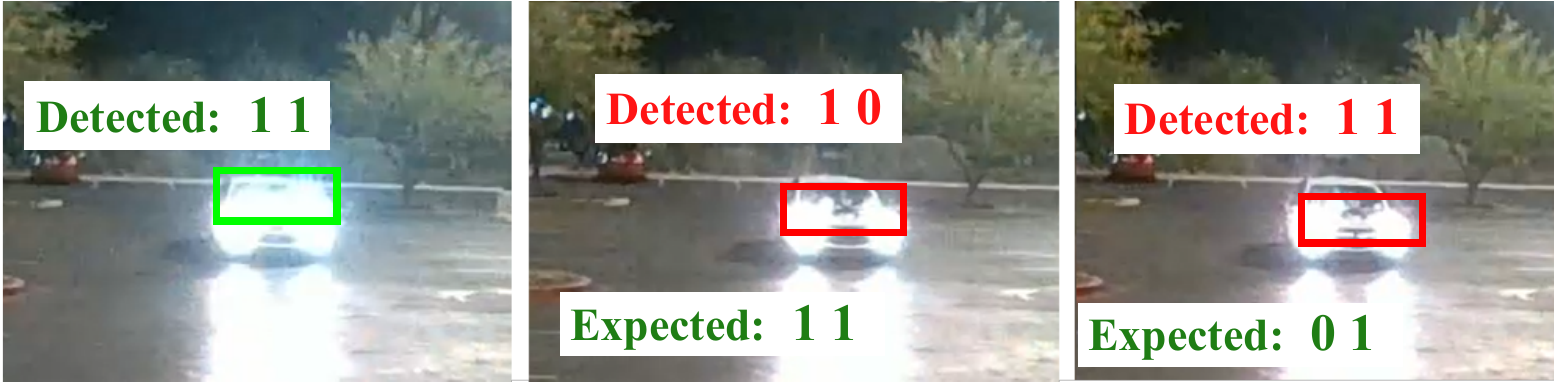} 
    \caption{Class 3 (11-11-01) misclassified as class 4 (11-10-11).}
    \label{fig:miss2}
  \end{subfigure}

  \caption{Misclassifications examples. Note: The patterns are flashed from the vehicle's perspective. Therefore, binary values such as 10 and 01 are interpreted by the camera in reverse (i.e., 10 is seen as 01 and vice versa).}
  \label{fig:misclassification}
\end{figure*}

\begin{figure}[t]
    \centering
    \includegraphics[width=0.98\linewidth]{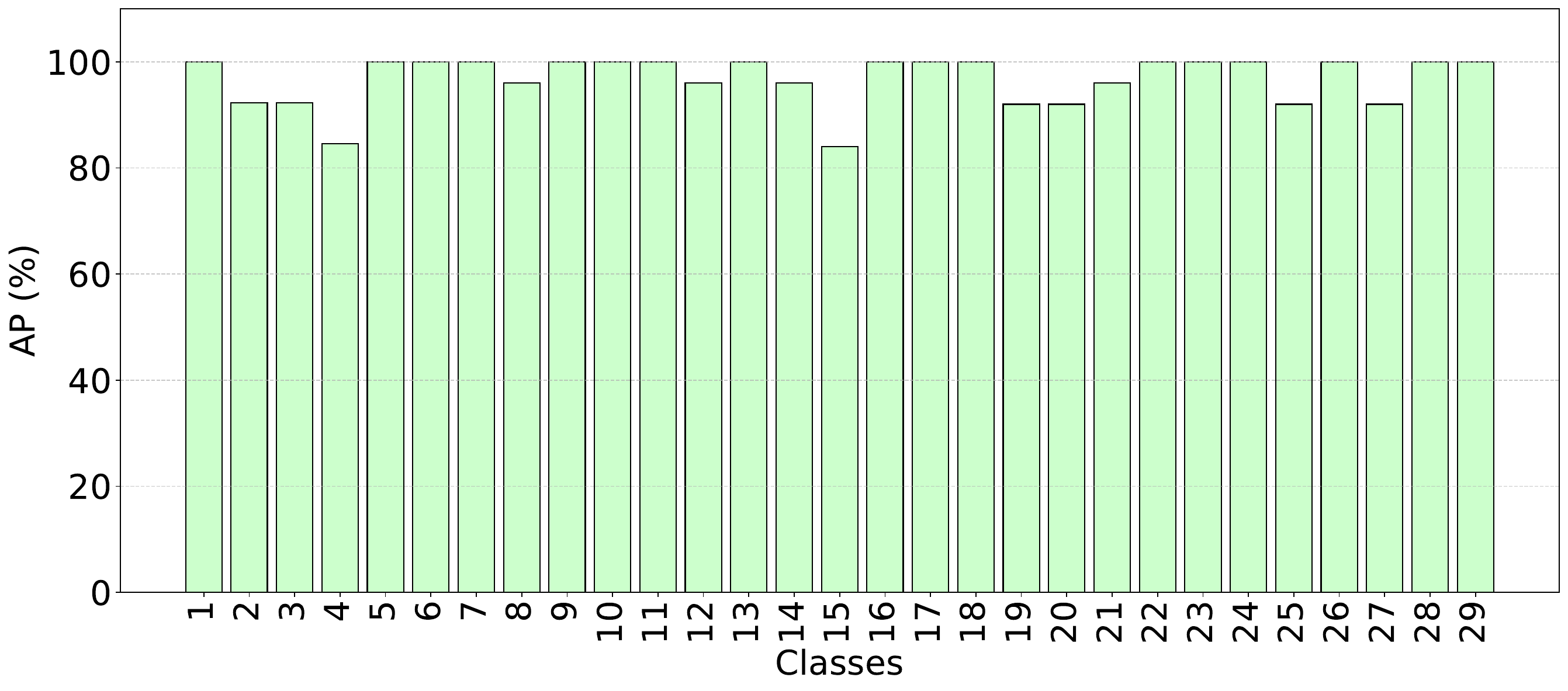}
    \caption{Per-class Average Precision (AP) across 29 classes in the real-car testbed.}
    \label{fig:per_class_ap}
\end{figure}

A detailed analysis of each testbed reveals several similarities in misclassifications. 
Incorrect predictions often occur between classes that differ only slightly in flash patterns or share similar temporal structures. As shown in \cref{fig:misclassification}, for example, in the RC-car testbed some sequences labeled as class 15 (10-10-01 were misclassified as class 14 (10-10-10). Similarly, in the real-car testbed, sequences from class 3 (11-11-01) were misclassified as class 4 (11-10-11). This confusion typically occurs when headlight flashing patterns significantly overlap, underscoring the challenge of distinguishing closely related classes. Nevertheless, the misclassification rate remains low, demonstrating that the proposed SlowFast-based model effectively captures the essential spatio-temporal features of headlight flashes.

A relevant finding from our experiments is the difference in inference speed between the two testbeds. On average, each video in the RC-car testbed is processed in about 1 ms. Although occasional outliers exceed this time, they have little effect on overall performance. The real-world real-car testbed also shows significant gains over the previous \ac{CNN} model, with an average processing time of 3.8 ms and most videos completing in just a few milliseconds. Even when inference time reaches several tenths of a second in rare cases, system performance remains stable. These fast inference times make the system suitable for near-real-time applications. This efficiency stems from the dual-path design of the SlowFast architecture. The fast path captures quick motion by sampling frames at high temporal resolution, while the slow path extracts broader context from frames at lower rates. This balance enables the model to detect both short- and long-term patterns without high computational cost. 

As shown in Figure~\ref{fig:comparison_plot}, SlowFast significantly outperforms Dwyer’s model in sequence detection latency, achieving lower and more consistent times. While Dwyer’s model shows a wider latency range, especially for vehicle detection (75–140 ms), it performs competitively in image cropping. However, its sequence detection latency is higher. The 3D \ac{CNN} shows low latency but supports fewer applications. The figure highlights how our real-car implementation of SlowFast maintains sub-5 ms performance with minimal variation, proving its suitability for time-critical vehicular tasks. Summing all three stages, vehicle detection, cropping, and sequence detection, Dwyer’s model averages 75–100 ms, far exceeding SlowFast (1.6 ms) and the 3D \ac{CNN} (just over 1 ms). This clear gap reinforces the value of our architecture for real-world, near-real-time deployment.

\begin{figure}[t!]
\centering
\includegraphics[width=0.45\textwidth]{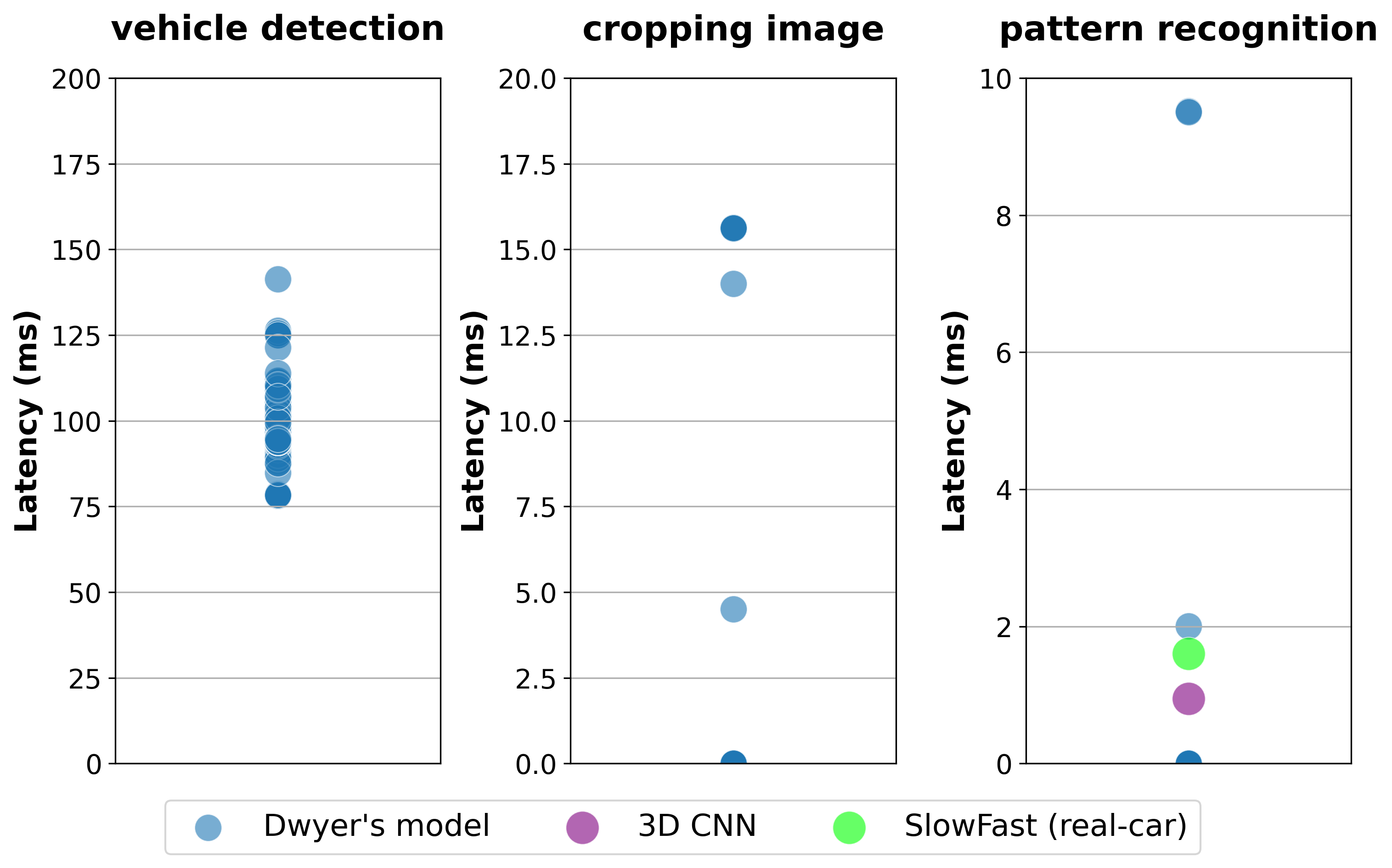}
\caption{Comparison of message decoding latencies between the different models.}
\label{fig:comparison_plot}
\end{figure}

%\begin{comment}
Finally, we present a summarized comparison of the two evaluated networks, 3D CNN and SlowFast, as shown in the \cref{tab:nn_architecture_comparison}.

\begin{table*}[htbp]
    \caption{Comparison of Neural Network Architectures for Headlight Flash Classification}
    \label{tab:nn_architecture_comparison}
    \renewcommand{\arraystretch}{1.15}
    \begin{center}
    \begin{small}
    \begin{tabular}{|l|p{6.5cm}|p{6.5cm}|}
        \hline
        \textbf{Component} & \textbf{3D CNN} & \textbf{SlowFast CNN} \\
        \hline
        \textbf{Input} & 
        Video clips resized to uniform dimensions. Each clip is processed as a full spatiotemporal tensor. &
        32 frames extracted from video. Fast pathway receives full frame rate; slow pathway receives every 4th frame ($\alpha$= 4). \\
        \hline
        \textbf{Architecture} & 
        3 convolutional blocks with:
        \begin{itemize}
            \item 3D Conv ($3\times3\times3$), padding=1
            \item ReLU
            \item MaxPooling ($2\times2\times2$)
        \end{itemize}
        Followed by:
        \begin{itemize}
            \item Dense layer (512 units, ReLU)
            \item Output layer (28 classes)
        \end{itemize}
        & 
        Dual-pathway ResNet-50:
        \begin{itemize}
            \item Slow Pathway: lower temporal resolution, high spatial
            \item Fast Pathway: higher temporal resolution, low channel depth
        \end{itemize}
        Lateral fusion from Fast → Slow. Final FC layer adapted to 28 or 29 classes. \\
        \hline
        \textbf{Pretraining} & None & Pretrained on Kinetics-400 \\
        \hline
        \textbf{Dropout} & Not used & Dropout before final layer (rate = 0.2) \\
        \hline
        \textbf{Optimizer} & AdamW & Adam \\
        \hline
        \textbf{Learning Rate} & $5\times10^{-4}$ & $1\times10^{-4}$ (with warm-up and cosine annealing) \\
        \hline
        \textbf{Batch Size} & 8 & 4 \\
        \hline
        \textbf{Epochs} & 15 & 32 \\
        \hline
        \textbf{Scheduler} & Linear scheduler & 2-epoch warm-up + cosine annealing \\
        \hline
        \textbf{Initialization} & Xavier initialization & Pretrained weights + FC layer initialized randomly \\
        \hline
        \textbf{Output Classes} & 28 classes (RC-car) & 28 (RC-car) or 29 (real-car, includes random flash class) \\
        \hline
    \end{tabular}
    \end{small}
    \end{center}
\end{table*}
%\end{comment}

\subsection{Handling multiple vehicles and high traffic conditions}\label{sec:handlingMul}

To manage scenarios in which multiple vehicles simultaneously flash their headlights, whether for authentication or other purposes, we propose introducing a preliminary system for spatial tracking and Region of Interest (ROI) extraction before classification. This system detects each vehicle within the camera field of view and isolates the relevant regions showing headlight activity. Specifically, an object detector with a Multi-Object Tracking (MOT) algorithm (for example, an object detector YOLO \cite{7780460} with a MOT tracker like ByteTrack~\cite{zhang2022bytetrack}) is used to identify bounding boxes for each detected vehicle. Once a bounding box is established, the system monitors the vehicle’s headlights over time to confirm if it is actively flashing.

After identifying vehicles that are flashing, we generate a subclip or spatio-temporal ROI for each bounding box. Each subclip captures only the frames and spatial regions corresponding to that vehicle's headlights, effectively filtering out interference from nearby vehicles. This approach ensures strong isolation, as each instance of our classification model, e.g., the SlowFast network, processes only one vehicle at a time, significantly reducing noise and ambiguity. In this way, the entire system forms a two-stage pipeline, which first detects and identifies the ROI sequences for each flashing vehicle, and then applies a video classification network to recognize the LOS channels.

\section{Conclusion and future work} \label{sec:conclusion} 
In this work, we propose a unified \ac{NLOS} and \ac{LOS} \ac{MFA} solution for vehicle authentication, particularly for applications such as signal preemption at intersections and
lane access control.
Our security scheme is designed to mitigate remote and local attacks.
Using a camera and a \ac{CNN} model for the classification of flashing headlight sequences enables efficient real-time pattern recognition. 
By integrating \ac{LOS} and \ac{NLOS} channels into a unified \ac{MFA} framework, our approach increases the security of \ac{V2I} communications without imposing computational overhead. 
Unlike existing solutions that rely solely on cryptographic authentication, our approach relies on an additional layer of physical verification, making remote spoofing and unauthorized access harder. 
These results demonstrate that our approach is conceptually robust and experimentally validated.

To further develop our current approach, several challenges must be addressed. Firstly, our scheme should be formalized into a protocol. The communication protocol between the vehicle and the road infrastructure must be designed, defined, and then tested for security. Secondly, a key focus will be on scalability in traffic scenarios, where multiple vehicles may seek authentication simultaneously. This requires optimizing system performance to handle real-world conditions efficiently. To address scalability in high traffic scenarios (as described in Section~\ref{sec:handlingMul}), our approach can incorporate spatial tracking and ROI extraction to ensure that each flashing vehicle is processed independently, maintaining system performance even when multiple vehicles request authentication simultaneously.

In addition, real-world road experimentation is essential to validate the performance of the system under various weather and environmental conditions, which are currently constrained by hardware limitations in our test environment. Improving sensor capabilities and hardware adaptability will be crucial for broader applicability. Beyond urban intersections, our aim is to extend our solution to other sensitive authentication scenarios, such as \ac{AV} access to airports, military zones, and dedicated lanes. These environments require heightened security and stricter authentication mechanisms, making them ideal testbeds to evaluate the robustness of our approach.

By tackling these challenges, we aim to increase the scalability, adaptability, and security of our solution, paving the way for a broader deployment in \ac{ITS}.

\bibliographystyle{IEEEtranN}
\bibliography{MAIN}

% Generated by IEEEtranN.bst, version: 1.14 (2015/08/26)
\begin{thebibliography}{34}
\providecommand{\natexlab}[1]{#1}
\providecommand{\url}[1]{#1}
\csname url@samestyle\endcsname
\providecommand{\newblock}{\relax}
\providecommand{\bibinfo}[2]{#2}
\providecommand{\BIBentrySTDinterwordspacing}{\spaceskip=0pt\relax}
\providecommand{\BIBentryALTinterwordstretchfactor}{4}
\providecommand{\BIBentryALTinterwordspacing}{\spaceskip=\fontdimen2\font plus
\BIBentryALTinterwordstretchfactor\fontdimen3\font minus \fontdimen4\font\relax}
\providecommand{\BIBforeignlanguage}[2]{{%
\expandafter\ifx\csname l@#1\endcsname\relax
\typeout{** WARNING: IEEEtranN.bst: No hyphenation pattern has been}%
\typeout{** loaded for the language `#1'. Using the pattern for}%
\typeout{** the default language instead.}%
\else
\language=\csname l@#1\endcsname
\fi
#2}}
\providecommand{\BIBdecl}{\relax}
\BIBdecl

\bibitem[Bodei et~al.(2023)Bodei, {De Vincenzi}, and Matteucci]{sdv}
C.~Bodei, M.~{De Vincenzi}, and I.~Matteucci, ``From hardware-functional to software-defined vehicles and their security issues,'' in \emph{2023 IEEE 21st International Conference on Industrial Informatics (INDIN)}, 2023, pp. 1--10.

\bibitem[{De Vincenzi} et~al.(2024){De Vincenzi}, Pesé, Bodei, Matteucci, Brooks, Hasan, Saracino, Hamad, and Steinhorst]{devincenzi2024contextualizingsecurityprivacysoftwaredefined}
\BIBentryALTinterwordspacing
M.~{De Vincenzi}, M.~D. Pesé, C.~Bodei, I.~Matteucci, R.~R. Brooks, M.~Hasan, A.~Saracino, M.~Hamad, and S.~Steinhorst, ``Contextualizing security and privacy of software-defined vehicles: State of the art and industry perspectives,'' 2024. [Online]. Available: \url{https://arxiv.org/abs/2411.10612}
\BIBentrySTDinterwordspacing

\bibitem[Darbha et~al.(2019)Darbha, Konduri, and Pagilla]{8463512}
S.~Darbha, S.~Konduri, and P.~R. Pagilla, ``{Benefits of V2V Communication for Autonomous and Connected Vehicles},'' \emph{IEEE Transactions on Intelligent Transportation Systems}, vol.~20, no.~5, pp. 1954--1963, 2019.

\bibitem[Rahman et~al.(2024)Rahman, Islam, Ball, and Goodin]{10.1117/12.3013514}
\BIBentryALTinterwordspacing
M.~Rahman, F.~Islam, J.~E. Ball, and C.~Goodin, ``{Traffic light recognition and V2I communications of an autonomous vehicle with the traffic light for effective intersection navigation using {YOLOv8} and {MAVS} simulation},'' in \emph{Autonomous Systems: Sensors, Processing, and Security for Ground, Air, Sea, and Space Vehicles and Infrastructure 2024}, M.~C. Dudzik, S.~M. Jameson, and T.~J. Axenson, Eds., vol. 13052, International Society for Optics and Photonics.\hskip 1em plus 0.5em minus 0.4em\relax SPIE, 2024, p. 130520J. [Online]. Available: \url{https://doi.org/10.1117/12.3013514}
\BIBentrySTDinterwordspacing

\bibitem[Abu-Nimeh(2011)]{Abu-Nimeh2011}
\BIBentryALTinterwordspacing
S.~Abu-Nimeh, \emph{Three-Factor Authentication}.\hskip 1em plus 0.5em minus 0.4em\relax Boston, MA: Springer US, 2011, pp. 1287--1288. [Online]. Available: \url{https://doi.org/10.1007/978-1-4419-5906-5_793}
\BIBentrySTDinterwordspacing

\bibitem[De~Vincenzi et~al.(2025)De~Vincenzi, Moore, Smith, Sarma, and Matteucci]{10818588}
M.~De~Vincenzi, J.~Moore, B.~Smith, S.~E. Sarma, and I.~Matteucci, ``Security risks and designs in the connected vehicle ecosystem: In-vehicle and edge platforms,'' \emph{IEEE Open Journal of Vehicular Technology}, vol.~6, pp. 442--454, 2025.

\bibitem[Dasgupta et~al.(2022)Dasgupta, Hollis, Rahman, Atkison, and Jones]{Dasgupta2022}
\BIBentryALTinterwordspacing
S.~Dasgupta, C.~Hollis, M.~Rahman, T.~Atkison, and S.~Jones, ``An innovative attack modeling and attack detection approach for a waiting time-based adaptive traffic signal controller,'' in \emph{International Conference on Transportation and Development 2022}.\hskip 1em plus 0.5em minus 0.4em\relax American Society of Civil Engineers, Aug. 2022, p. 72–84. [Online]. Available: \url{http://dx.doi.org/10.1061/9780784484326.008}
\BIBentrySTDinterwordspacing

\bibitem[Irfan et~al.(2022)Irfan, Rahman, Atkison, Dasgupta, and Hainen]{irfan2022reinforcementlearningbasedcyberattack}
\BIBentryALTinterwordspacing
M.~S. Irfan, M.~Rahman, T.~Atkison, S.~Dasgupta, and A.~Hainen, ``Reinforcement learning based cyberattack model for adaptive traffic signal controller in connected transportation systems,'' 2022. [Online]. Available: \url{https://arxiv.org/abs/2211.01845}
\BIBentrySTDinterwordspacing

\bibitem[Yu et~al.(2022)Yu, Bai, Luan, and Qi]{9916277}
W.~Yu, W.~Bai, W.~Luan, and L.~Qi, ``State-of-the-art review on traffic control strategies for emergency vehicles,'' \emph{IEEE Access}, vol.~10, pp. 109\,729--109\,742, 2022.

\bibitem[Kanthavel et~al.(2021)Kanthavel, Sangeetha, and Keerthana]{V2I1}
\BIBentryALTinterwordspacing
D.~Kanthavel, S.~Sangeetha, and K.~Keerthana, ``An empirical study of vehicle to infrastructure communications - an intense learning of smart infrastructure for safety and mobility,'' \emph{International Journal of Intelligent Networks}, vol.~2, pp. 77--82, 2021, doi: \url{10.1016/j.ijin.2021.06.003}. [Online]. Available: \url{https://www.sciencedirect.com/science/article/pii/S2666603021000105}
\BIBentrySTDinterwordspacing

\bibitem[Yoshizawa et~al.(2023)Yoshizawa, Singel{\'{e}}e, M{\"{u}}hlberg, Delbruel, Taherkordi, Hughes, and Preneel]{surveyV2Xattack2}
\BIBentryALTinterwordspacing
T.~Yoshizawa, D.~Singel{\'{e}}e, J.~T. M{\"{u}}hlberg, S.~Delbruel, A.~Taherkordi, D.~Hughes, and B.~Preneel, ``A survey of security and privacy issues in {V2X} communication systems,'' \emph{{ACM} Comput. Surv.}, vol.~55, no.~9, pp. 185:1--185:36, 2023. [Online]. Available: \url{https://doi.org/10.1145/3558052}
\BIBentrySTDinterwordspacing

\bibitem[Qian et~al.(2022)Qian, Wang, Yang, and Xu]{surveyV2Xattack3}
\BIBentryALTinterwordspacing
J.~Qian, W.~Wang, X.~Yang, and H.~Xu, ``Survey on security and privacy in {5G} {V2X},'' in \emph{2022 6th International Conference on Electronic Information Technology and Computer Engineering, {EITCE} 2022, Xiamen, China, October 21-23, 2022}.\hskip 1em plus 0.5em minus 0.4em\relax {ACM}, 2022, pp. 1056--1062. [Online]. Available: \url{https://doi.org/10.1145/3573428.3573618}
\BIBentrySTDinterwordspacing

\bibitem[Alsoliman et~al.(2021)Alsoliman, Levorato, and Chen]{visual1}
A.~Alsoliman, M.~Levorato, and Q.~A. Chen, ``Vision-based two-factor authentication and localization scheme for autonomous vehicles,'' in \emph{Third International Workshop on Automotive and Autonomous Vehicle Security (AutoSec) 2021 (part of NDSS)}, 2021, doi: \url{10.14722/autosec.2021.23021}.

\bibitem[De~Vincenzi et~al.(2024)De~Vincenzi, Bodei, and Matteucci]{10.1145/3605098.3636102}
\BIBentryALTinterwordspacing
M.~De~Vincenzi, C.~Bodei, and I.~Matteucci, ``Olive: Flexible, portable, and sustainable {V2X}multi-factor authentication,'' in \emph{39th ACM/SIGAPP Symposium on Applied Computing}, ser. SAC '24.\hskip 1em plus 0.5em minus 0.4em\relax New York, NY, USA: Association for Computing Machinery, 2024, p. 215–217. [Online]. Available: \url{https://doi.org/10.1145/3605098.3636102}
\BIBentrySTDinterwordspacing

\bibitem[Song et~al.(2022)Song, Liu, Zhang, Yang, and Fu]{taillight}
\BIBentryALTinterwordspacing
W.~Song, S.~Liu, T.~Zhang, Y.~Yang, and M.~Fu, ``Action-state joint learning-based vehicle taillight recognition in diverse actual traffic scenes,'' \emph{{IEEE} Trans. Intell. Transp. Syst.}, vol.~23, no.~10, pp. 18\,088--18\,099, 2022. [Online]. Available: \url{https://doi.org/10.1109/TITS.2022.3160501}
\BIBentrySTDinterwordspacing

\bibitem[Suo and Sarma(2022)]{dajiangMFA}
D.~Suo and S.~E. Sarma, ``A two-factor authentication scheme for moving connected vehicles,'' in \emph{2022 IEEE 96th Vehicular Technology Conference (VTC2022-Fall)}.\hskip 1em plus 0.5em minus 0.4em\relax IEEE, 2022, pp. 1--5, doi: \url{10.1109/VTC2022-Fall57202.2022.10012773}.

\bibitem[Dwyer et~al.(2023)Dwyer, Sarma, and Suo]{benDw}
B.~Dwyer, S.~E. Sarma, and D.~Suo, ``Enabling secure vehicle to infrastructure communication via two-factor authentication,'' in \emph{IEEE 26th International Conference on Intelligent Transportation Systems (ITSC)}, 2023, pp. 5663--5668, doi: \url{10.1109/ITSC57777.2023.10421946}.

\bibitem[Redmon et~al.(2016{\natexlab{a}})Redmon, Divvala, Girshick, and Farhadi]{redmon2016you}
J.~Redmon, S.~Divvala, R.~Girshick, and A.~Farhadi, ``You only look once: Unified, real-time object detection,'' in \emph{2016 IEEE Conference on Computer Vision and Pattern Recognition (CVPR)}, 2016, pp. 779--788, doi: \url{10.1109/CVPR.2016.91}.

\bibitem[Arfaoui et~al.(2020)Arfaoui, Soltani, Tavakkolnia, Ghrayeb, Safari, Assi, and Haas]{Arfaoui}
M.~A. Arfaoui, M.~D. Soltani, I.~Tavakkolnia, A.~Ghrayeb, M.~Safari, C.~M. Assi, and H.~Haas, ``Physical layer security for visible light communication systems: A survey,'' \emph{IEEE Communications Surveys and Tutorials}, vol.~22, no.~3, pp. 1887--1908, 2020.

\bibitem[Singh et~al.(2022)Singh, Srivastava, Bohara, Liu, Noor-A-Rahim, and Ghatak]{Singh}
G.~Singh, A.~Srivastava, V.~A. Bohara, Z.~Liu, M.~Noor-A-Rahim, and G.~Ghatak, ``Heterogeneous visible light and radio communication for improving safety message dissemination at road intersection,'' \emph{IEEE Transactions on Intelligent Transportation Systems}, vol.~23, no.~10, pp. 17\,607--17\,619, 2022.

\bibitem[Rowan et~al.(2017)Rowan, Clear, Gerla, Huggard, and Mc~Goldrick]{Rowan}
S.~Rowan, M.~Clear, M.~Gerla, M.~Huggard, and C.~Mc~Goldrick, ``Securing vehicle to vehicle communications using blockchain through visible light and acoustic side-channels,'' 04 2017.

\bibitem[Shaaban and Faruque(2020)]{Shaaban}
\BIBentryALTinterwordspacing
R.~Shaaban and S.~Faruque, ``Cyber security vulnerabilities for outdoor vehicular visible light communication in secure platoon network: Review, power distribution, and signal to noise ratio analysis,'' \emph{Physical Communication}, vol.~40, p. 101094, 2020. [Online]. Available: \url{https://www.sciencedirect.com/science/article/pii/S1874490720301701}
\BIBentrySTDinterwordspacing

\bibitem[5GAA()]{5gaaUS}
\BIBentryALTinterwordspacing
5GAA. {United States Vehicle-to-Infrastructure Communications; Day One Deployment Guide}. Last accessed on March 10, 2025. [Online]. Available: \url{https://5gaa.org/content/uploads/2023/10/5gaa-wi-usdploy-231667-technical-report-guidance-day-1.pdf}
\BIBentrySTDinterwordspacing

\bibitem[OWASP(2023)]{stride}
\BIBentryALTinterwordspacing
OWASP. (2023) {STRIDE} model. Last accessed on March 10, 2025. [Online]. Available: \url{https://owasp.org/www-community/Threat_Modeling_Process}
\BIBentrySTDinterwordspacing

\bibitem[Dolev and Yao(1983)]{dolevYao}
\BIBentryALTinterwordspacing
D.~Dolev and A.~C. Yao, ``On the security of public key protocols,'' \emph{{IEEE} Trans. Inf. Theory}, vol.~29, no.~2, pp. 198--207, 1983. [Online]. Available: \url{https://doi.org/10.1109/TIT.1983.1056650}
\BIBentrySTDinterwordspacing

\bibitem[{CyberNews - Gintaras Radauskas}(2024)]{cybernews2024dutchtrafficlights}
\BIBentryALTinterwordspacing
{CyberNews - Gintaras Radauskas}. (2024) Dutch government to replace hackable traffic lights. Last accessed on March 10, 2025. [Online]. Available: \url{https://cybernews.com/news/dutch-government-will-replace-hackable-traffic-lights/}
\BIBentrySTDinterwordspacing

\bibitem[iee()]{ieee16092}
\BIBentryALTinterwordspacing
{IEEE 1609.2 Standard for Wireless Access in Vehicular Environments}. Last accessed on March 1st, 2025. [Online]. Available: \url{https://ieeexplore.ieee.org/stamp/stamp.jsp?arnumber=7426684}
\BIBentrySTDinterwordspacing

\bibitem[{IEEE Standards}(2023)]{10075082}
{IEEE Standards}, ``{IEEE} approved draft standard for wireless access in vehicular environments--security services for applications and management messages,'' \emph{IEEE Std 1609.2-2022 (Revision of IEEE Std 1609.2-2016)}, pp. 1--349, 2023.

\bibitem[Brecht et~al.(2018)Brecht, Therriault, Weimerskirch, Whyte, Kumar, Hehn, and Goudy]{brecht2018security}
B.~Brecht, D.~Therriault, A.~Weimerskirch, W.~Whyte, V.~Kumar, T.~Hehn, and R.~Goudy, ``A security credential management system for {V2X} communications,'' \emph{IEEE Transactions on Intelligent Transportation Systems}, vol.~19, no.~12, pp. 3850--3871, 2018, doi: \url{10.1109/TITS.2018.2797529}.

\bibitem[Feichtenhofer et~al.(2018)Feichtenhofer, Fan, Malik, and He]{slowFast}
\BIBentryALTinterwordspacing
C.~Feichtenhofer, H.~Fan, J.~Malik, and K.~He, ``Slowfast networks for video recognition,'' \emph{CoRR}, vol. abs/1812.03982, 2018. [Online]. Available: \url{http://arxiv.org/abs/1812.03982}
\BIBentrySTDinterwordspacing

\bibitem[PyTorch(2021)]{pytorchvideo_slowfast}
PyTorch, ``Slowfast video models for {PyTorch},'' \url{https://pytorch.org/hub/facebookresearch_pytorchvideo_slowfast/}, 2021, last accessed on March 10, 2025.

\bibitem[Kay et~al.(2017)Kay, Carreira, Simonyan, Zhang, Hillier, Vijayanarasimhan, Viola, Green, Back, Natsev, Suleyman, and Zisserman]{kay2017kineticshumanactionvideo}
\BIBentryALTinterwordspacing
W.~Kay, J.~Carreira, K.~Simonyan, B.~Zhang, C.~Hillier, S.~Vijayanarasimhan, F.~Viola, T.~Green, T.~Back, P.~Natsev, M.~Suleyman, and A.~Zisserman, ``The kinetics human action video dataset,'' 2017. [Online]. Available: \url{https://arxiv.org/abs/1705.06950}
\BIBentrySTDinterwordspacing

\bibitem[Redmon et~al.(2016{\natexlab{b}})Redmon, Divvala, Girshick, and Farhadi]{7780460}
J.~Redmon, S.~Divvala, R.~Girshick, and A.~Farhadi, ``You only look once: Unified, real-time object detection,'' in \emph{2016 IEEE Conference on Computer Vision and Pattern Recognition (CVPR)}, 2016, pp. 779--788.

\bibitem[Zhang et~al.(2022)Zhang, Sun, Jiang, Yu, Weng, Yuan, Luo, Liu, and Wang]{zhang2022bytetrack}
Y.~Zhang, P.~Sun, Y.~Jiang, D.~Yu, F.~Weng, Z.~Yuan, P.~Luo, W.~Liu, and X.~Wang, ``Bytetrack: Multi-object tracking by associating every detection box,'' in \emph{European conference on computer vision}.\hskip 1em plus 0.5em minus 0.4em\relax Springer, 2022.

\end{thebibliography}

\begin{IEEEbiography}
[{\includegraphics[width=1in,height=1in,clip,keepaspectratio]{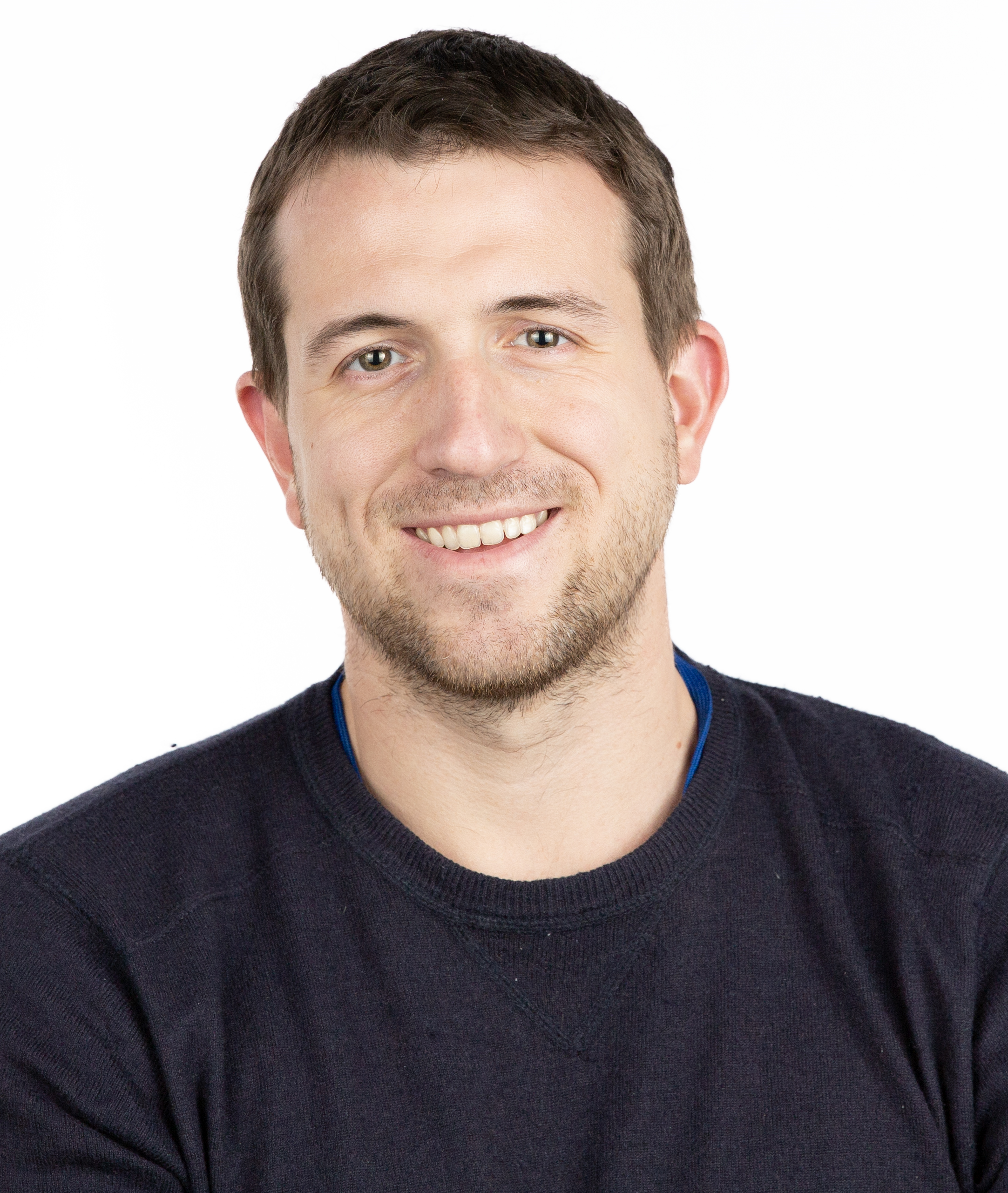}}]{Marco De Vincenzi}~is a researcher at the Italian National Research Center (CNR). He was a visiting professor at the AutoID Lab at the Massachusetts Institute of Technology and at Prof. Suo's lab at Arizona State University. He earned a Master of Computer Science in Data Science and Business Informatics and a Master in Cybersecurity. He spent six years in the automotive industry. His research interests include security and privacy in automotive, in particular authentication processes. 
\end{IEEEbiography}

\begin{IEEEbiography}[{\includegraphics[width=1in,height=1in,clip,keepaspectratio]{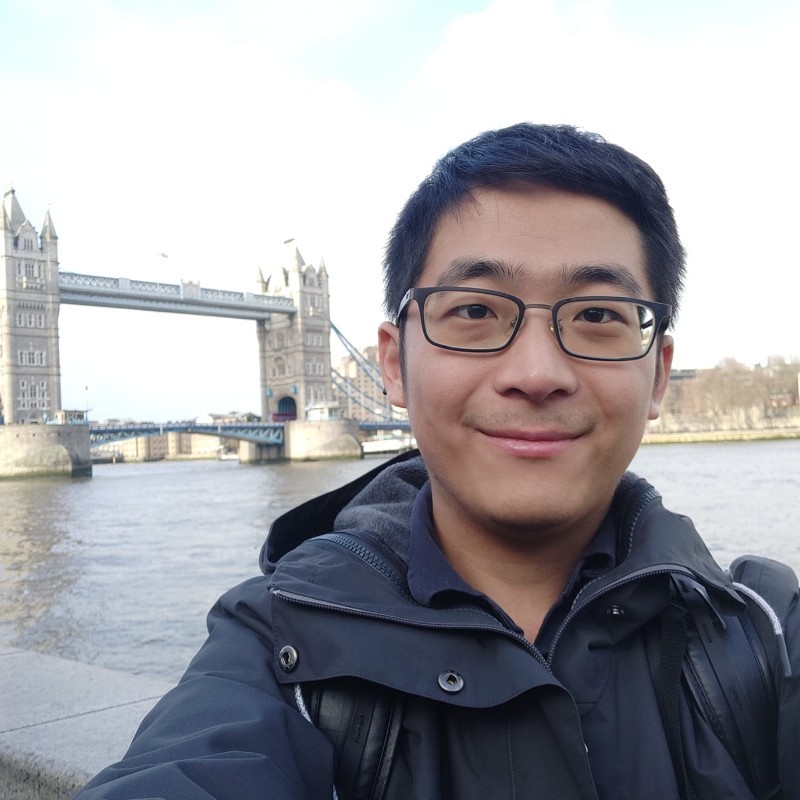}}] {Shuyang Sun} is a visiting fellow of Torr Vision Group, University of Oxford. He is also a research scientist at ByteDance, United States. He got his D.Phil. (Ph.D.) degree from University of Oxford in 2024. During his Ph.D., he also collaborated closely with researchers at Google DeepMind, Google Research, Intel ISL and ByteDance etc. Before that, Shuyang got his M.Phil. degree from the University of Sydney in 2019 and B.Eng. degree from Wuhan University in 2016. His research primarily focuses on computer vision and multi-modal learning.
\end{IEEEbiography}

\begin{IEEEbiography}[{\includegraphics[width=1in,height=1in,clip,keepaspectratio]{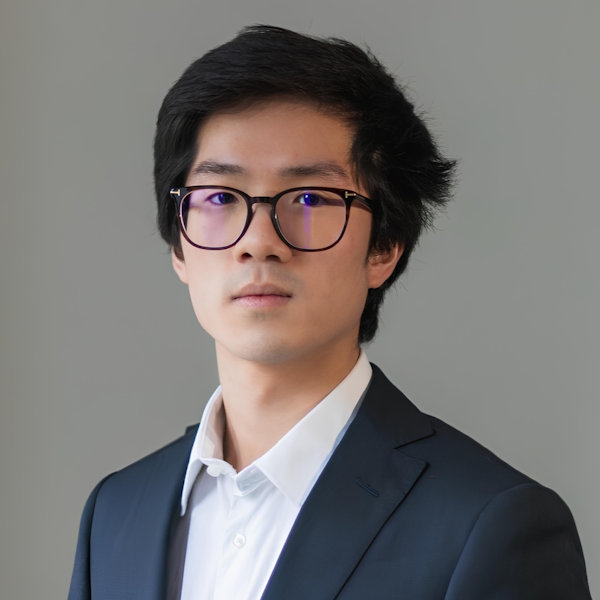}}] {Chen Bo Calvin Zhang} was a visiting researcher at the Massachusetts Institute of Technology (MIT). He is currently pursuing his Master of Science in Data Science at ETH Zurich. He previously earned his Bachelor of Science (Hons) in Computer Science and Mathematics from the University of Manchester. His research interests include reinforcement learning, sequential decision making, and AI alignment. During his academic career, he has focused on topics such as preference-based reinforcement learning and adversarial attacks for deep reinforcement learning.
\end{IEEEbiography}

\begin{IEEEbiography}[{\includegraphics[width=1in,height=1in,clip,keepaspectratio]{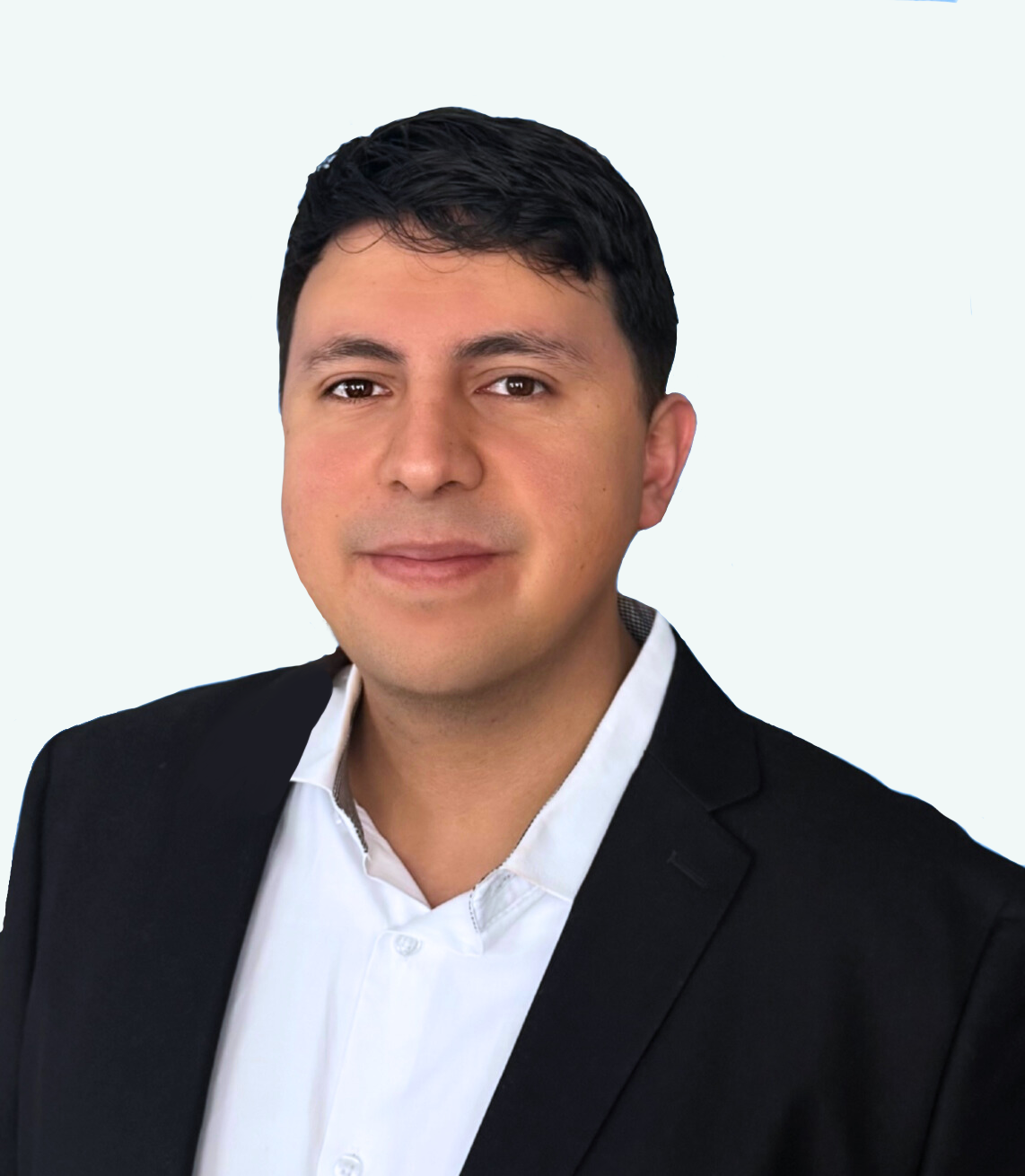}}] {Manuel Garcia Jr} is an undergraduate student at Arizona State University, currently pursuing a degree in Engineering (Electrical Systems). His primary focus is on Embedded Systems, specifically Microcontroller Processing and Digital Control. His research interests include MmWave Reflector Technology for advancements in drone localization and precision landing techniques. He was accepted as a Graduate Student in the Clean Energy Systems program at Arizona State University under the supervision of Prof. Suo.
\end{IEEEbiography}

\begin{IEEEbiography}[{\includegraphics[width=1in,height=1in,clip,keepaspectratio]{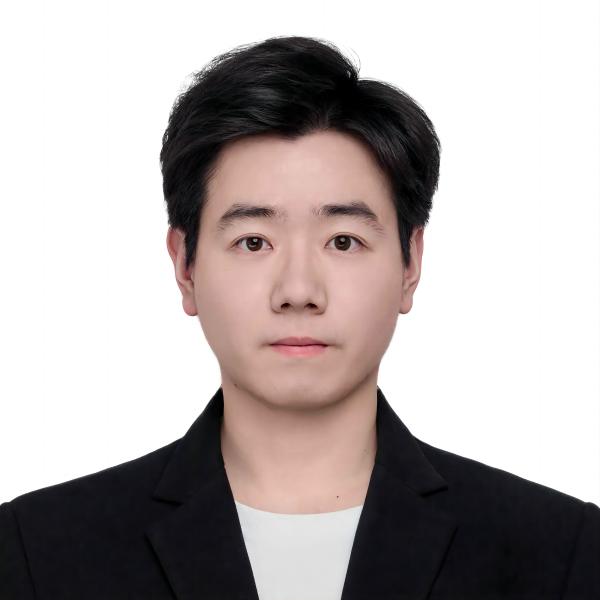}}] {Shaozu Ding} received the B.S. degree in control science and engineering, the M.S. degree in electronic information from Zhejiang University, Zhejiang, China, in 2021 and 2024, respectively. He is currently pursuing the Ph.D. degree in systems engineering with Arizona State University, Mesa, USA. His current research interests include digital twin, multi-modal sensor deployment optimization and multi-modal sensor 3D target detection.
\end{IEEEbiography}

\begin{IEEEbiography}
[{\includegraphics[width=1in,height=1in,clip,keepaspectratio]{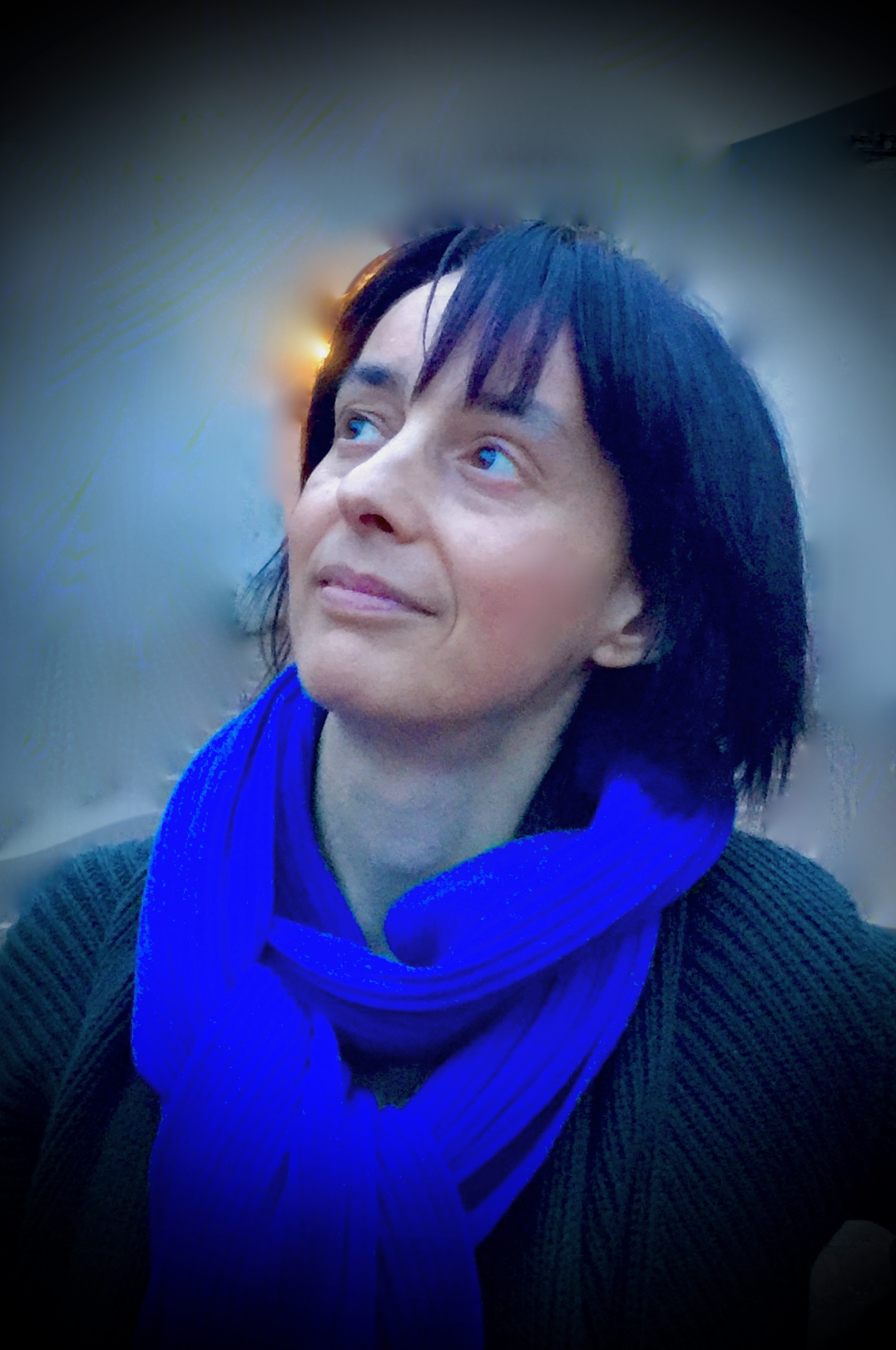}}]{Chiara Bodei}
(Ph.D. 2000) is an Associate Professor of Computer
Science since 2005 at the University of Pisa. Her research interests include the theory of concurrency and the security of distributed systems, networks, and the Internet of Things. In particular, her work focuses on the application of formal methods to the modeling and analysis of distributed systems, including IoT environments. To this aim, she has worked extensively with process algebras and Control Flow Analysis techniques. More recently, her research has turned to automotive cybersecurity, with particular attention to authentication processes. 
\end{IEEEbiography} 

\begin{IEEEbiography}
[{\includegraphics[width=1in,height=1in,clip,keepaspectratio]{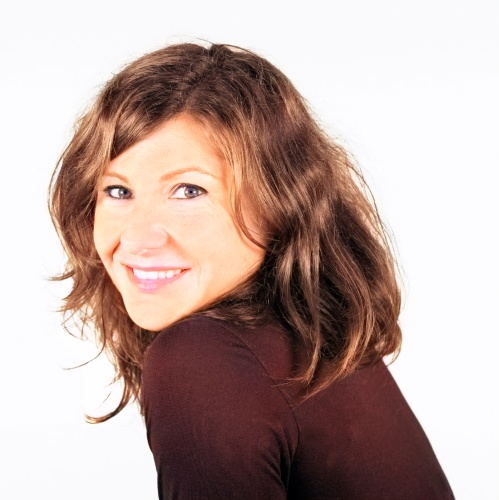}}]{Ilaria Matteucci}
 (M.Sc. 2003, Ph.D. 2008) is a researcher of the Trust, Security and Privacy group within the Institute of Informatics and Telematics of CNR. Her main research interests include formal methods for secure systems, analysis of data sharing and policies on personal data privacy. Currently, the research interest is focused on Automotive defensive and offensive cybersecurity. She participates in national and European projects in the field of information security.
\end{IEEEbiography}

\begin{IEEEbiography}[{\includegraphics[width=1in,height=1.25in,clip,keepaspectratio]{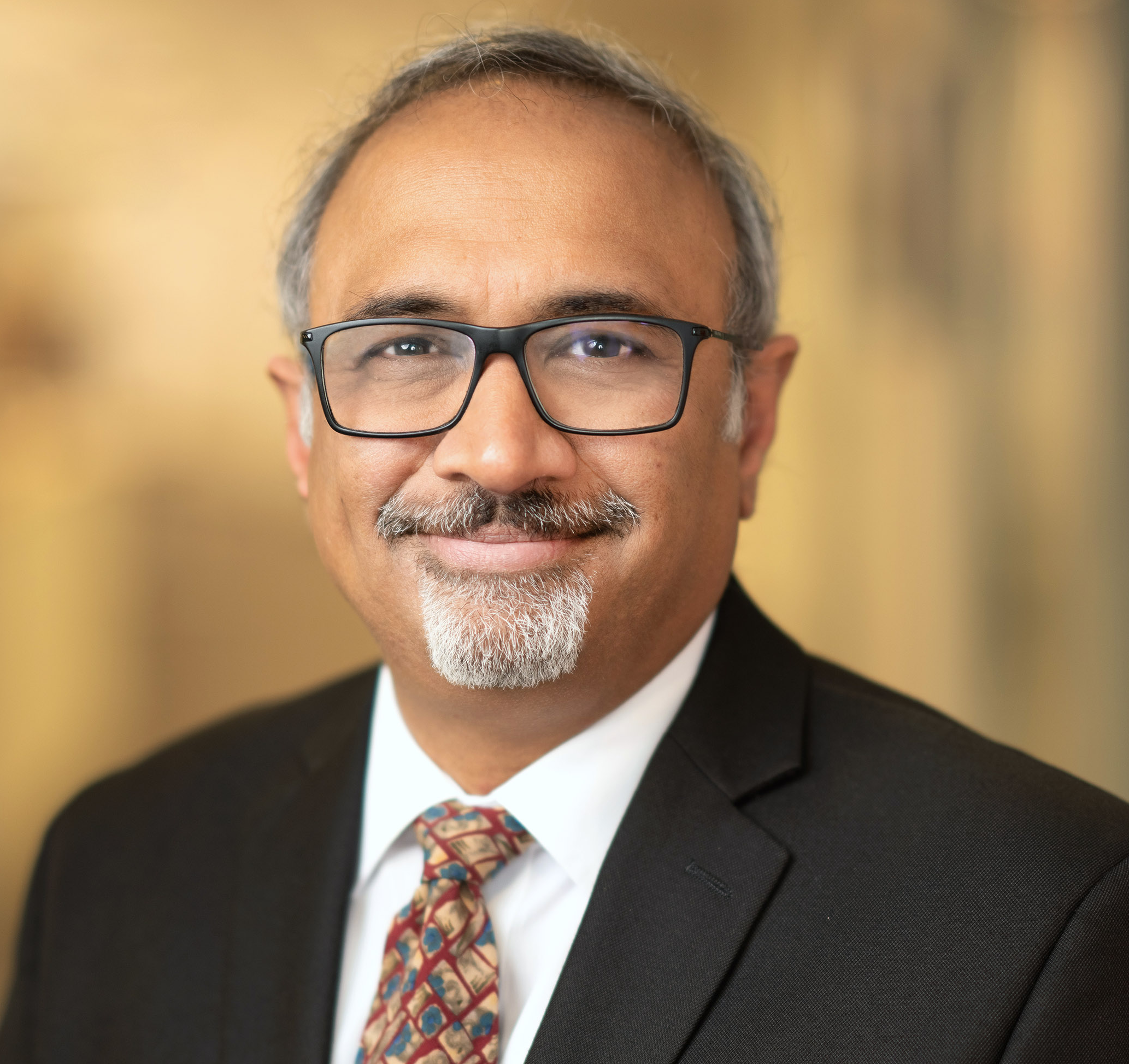}}]{Sanjay E. Sarma} is the Fred Fort Flowers (1941) and Daniel Fort Flowers (1941) Professor of Mechanical Engineering at MIT. He co-founded the Auto-ID Center at MIT and developed many of the key technologies behind the EPC suite of RFID standards now used worldwide. He was also the founder and CTO of OATSystems, which was acquired by Checkpoint Systems (NYSE: CKP) in 2008. He serves on the boards of GS1, EPCglobal and several companies including CleanLab and Aclara Resources (TSX:ARA). 

Prof. Sarma received his Bachelors from the Indian Institute of Technology, his Masters from Carnegie Mellon University and his PhD from the University of California at Berkeley. Sarma also worked at Schlumberger Oilfield Services in Aberdeen, UK. He has authored over 150 academic papers in computational geometry, sensing, RFID, automation and CAD, and is the recipient of numerous awards for teaching and research including the MacVicar Fellowship, the Business Week eBiz Award and Informationweek’s Innovators and Influencers Award.
\end{IEEEbiography}

\begin{IEEEbiography}[{\includegraphics[width=1in,height=1in,clip,keepaspectratio]{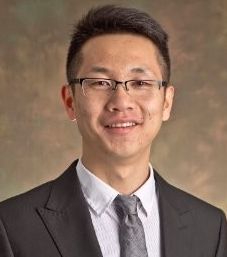}}] {Dajiang Suo} is an Assistant Professor at Arizona State University. He obtained a Ph.D. in Mechanical Engineering from MIT in 2020. Suo holds a B.S. degree in mechatronics engineering, and S.M. degrees in Computer Science and Engineering Systems. His research interests include secure connectivity (e.g., vehicle-to-everything communication) and multimodal sensing technologies for building cyber-resilient transportation systems. Before returning to school to pursue PhD degree, Suo was with the vehicle control and autonomous driving team at Ford Motor Company (Dearborn, MI), working on the safety and cyber-security of automated vehicles. He also serves as a paper editor for the Standing Committee on Enterprise, Systems, and Cyber Resilience (AMR40) at the Transportation Research Board.
\end{IEEEbiography}

\end{document}